\documentclass[12pt,a4paper]{article}

\usepackage{epsfig,multicol,multirow,cite}
\usepackage{amsmath,subfigure,latexsym,amssymb}

\def\lsi{\raise0.3ex\hbox{$<$\kern-0.75em\raise-1.1ex\hbox{$\sim$}}}
\def\gsi{\raise0.3ex\hbox{$>$\kern-0.75em\raise-1.1ex\hbox{$\sim$}}}
\def\backder{\raise1.4ex\hbox{$\leftarrow$\kern-0.75em\raise-1.4ex\hbox{$\partial$}}}
\def\forder{\raise1.4ex\hbox{$\rightarrow$\kern-0.75em\raise-1.4ex\hbox{$\partial$}}}

\newcommand{\lsim}{\mathop{\lsi}}
\newcommand{\gsim}{\mathop{\gsi}}

\newcommand{\be}{\begin{equation}}
\newcommand{\ee}{\end{equation}}
\newcommand{\nn}{\nonumber}
\newcommand{\bea}{\begin{eqnarray}}
\newcommand{\eea}{\end{eqnarray}}

\newcommand{\R}{{\kern+.25em\sf{R}\kern-.78em\sf{I} \kern+.78em\kern-.25em}}
\newcommand{\RR}{{\kern+.25em\sf{R}\kern-.6em\sf{I} \kern+.6em\kern-.25em}}
\newcommand{\N}{{\kern+.25em\sf{N}\kern-.78em\sf{I} \kern+.78em\kern-.25em}}
\newcommand{\C}{{\kern+.25em\sf{C}\kern-.50em\sf{I} \kern+.50em\kern-.25em}}

\newcommand{\upa}{\uparrow}
\newcommand{\doa}{\downarrow}

\makeatletter
\@addtoreset{equation}{section}
\makeatother

\begin{document}

\begin{center}

{\Large\bf Berezinskii-Kosterlitz-Thouless Transition}
\vspace*{6mm} \\
{\Large\bf with a Constraint Lattice Action}

\vspace*{10mm}

Wolfgang Bietenholz$^{\rm \, a}$, Urs Gerber$^{\rm \, a}$ \\
and Fernando G.\ Rej\'{o}n-Barrera$^{\rm \, b}$

\vspace*{6mm}

{\small
$^{\rm \, a}$  Instituto de Ciencias Nucleares \\
Universidad Nacional Aut\'{o}noma de M\'{e}xico \\
A.P. 70-543, C.P. 04510 Distrito Federal \\ 
Mexico

\vspace*{4mm}
$^{\rm \, b}$ Institute for Theoretical Physics \\
University of Amsterdam \\
Science Park 904 \\
Postbus 94485, 1090 GL Amsterdam \\ 
The Netherlands
}

\end{center}

\vspace*{4mm}

The 2d XY model exhibits an essential phase transition, which 
was predicted long ago --- by Berezinskii, Kosterlitz and Thouless
(BKT) --- to be driven by the (un)binding of vortex--anti-vortex pairs. 
This transition has been confirmed for the standard lattice action,
and for actions with distinct couplings, in agreement with
universality. Here we study a highly unconventional formulation of this 
model, which belongs to the class of topological lattice actions:
it does not have any couplings at all, but just a constraint
for the relative angles between nearest neighbour spins. By means of
dynamical boundary conditions we measure the helicity modulus $\Upsilon$, 
which shows that this formulation performs a BKT phase transition as
well. Its finite size effects are amazingly mild, in contrast to other 
lattice actions. This provides one of the most precise numerical
confirmations ever of a BKT transition in this model. On the other 
hand, up to the lattice sizes that we explored, there are deviations 
from the spin wave approximation, for instance for the Binder cumulant 
$U_{4}$ and for the leading finite size correction to $\Upsilon$.
Finally we observe that the (un)binding mechanism follows the usual 
pattern, although free vortices do not require any energy in this 
formulation. Due to that observation, one should reconsider an aspect 
of the established picture, which estimates the critical temperature 
based on this energy requirement.

\vspace*{6mm}

\noindent
{\footnotesize E-mail: \ wolbi@nucleares.unam.mx, 
\ gerber@correo.nucleares.unam.mx, \\
\hspace*{1.3cm} f.rejon@student.uva.nl}

\newpage

\tableofcontents

\section{The 2d XY model and its constraint lattice action}

The 2d XY model is one of the simplest models in quantum
field theory and statistical mechanics. 
It has been studied extensively since the 1970s, 
but interesting aspects are still being revealed.

This model describes certain systems in solid state physics, in particular
superfluid helium films \cite{Kos73}, which is reflected by the global 
$O(2)$ symmetry. Further applications include superconducting films
\cite{supracon}, the Coulomb gas model \cite{Froh},
Josephson junction arrays \cite{Jja} 
and nematic liquid crystals \cite{liqcry}.
The applications are not as broad as for the (even simpler) Ising model, 
but the phase transition in the XY model is conceptually more 
interesting. It is one of the few examples in the literature for a
transition beyond second order; more precisely it is of infinite order, 
and therefore an {\em essential phase transition,} known as the
Berezinskii-Kosterlitz-Thouless (BKT) transition \cite{Ber70,Kos73}.
BKT phase transitions have been identified also in other models,
which could be solved exactly \cite{exact}.

In the 2d XY model,
the key to its understanding are the {\em vortices and anti-vortices} 
\cite{Ber70,Kos73}. Entire configurations do not have distinct topological 
charges, but in a square lattice regularisation each plaquette
carries a winding number $0$, $+1$ (vortex) or $-1$ (anti-vortex). 
The dynamics of these topological defects turned out to be crucial 
for the phase diagram.

Here we stay within the framework of formulations on an $L \times L$ 
square lattice, with a classical spin $\vec e_{x}$
attached to each site $x$,
\be
\vec e_{x} = \left( \begin{array}{c} \cos \phi_{x} \\ \sin \phi_{x}
\end{array} \right) \in \R^{2} \ ,
\ee
such that $| \vec e_{x} | = 1\, , \ \forall x \,$.
The {\em standard action} on the lattice (in lattice units) reads
\bea
S [ \vec e \, ] &=& \sum_{\langle x y \rangle} 
s( \vec e_{x} , \vec e_{y} ) \ , \nn \\ 
s( \vec e_{x} , \vec e_{y} ) &=&
\beta \Big( 1 - \vec e_{x} \cdot \vec e_{y} \Big) 
= \beta \Big( 1 - \cos ( \phi_{x} - \phi_{y} ) \Big) \ ,
\label{stanact}
\eea
where the sum runs over all nearest neighbour sites $x$, $y$, 
and $\beta > 0$ is the inverse coupling.
Its critical value for the BKT transition was identified as 
$\beta_{\rm c} = 1.1199(1)$ \cite{HasPin} (earlier determinations
of $\beta_{\rm c}$ are quoted in Refs.\ \cite{KenIrv,Kenna}).

The mechanism behind the BKT phase transition was understood based 
on the density of free vortices (and anti-vortices):

\begin{itemize}

\item At $\beta > \beta_{\rm c}$ this density is low, so that a kind
of long-range order\footnote{We use the term ``order'' in a
general sense, beyond the specific meaning, which is
excluded in $d=2$ by the Mermin-Wagner Theorem.} emerges. 
The correlations only decay with a power law, 
so the correlation length $\xi$ is infinite {\em (massless phase).}
Most of the topological defects occur as tightly bound
vortex--anti-vortex pairs, which appear topologically neutral
from a large-scale perspective, hence these objects do not
prevent the long-range order.

\item As $\beta$ decreases below $\beta_{\rm c}$, a significant 
number of these pairs dissociate, so the density of free vortices jumps
up. This destroys the long-range order, and $\xi$ becomes finite
{\em (massive phase).} 
The value of $\beta_{\rm c}$ has been estimated from the energy that
a free vortex requires \cite{Kos73}.
\end{itemize}

This picture was originally in competition with other proposed
scenarios, but it is now generally accepted since it led to
correct quantitative predictions. They include the exponential
divergence of $\xi$ at $\beta \lsim \beta_{\rm c}$ as \cite{Kos74}
\be  \label{nue}
\xi \propto \exp \Big( \frac{\rm const.}{(\beta_{\rm c} 
- \beta )^{\nu}} \Big) \ , \qquad \nu_{\rm c} = 1/2 \ ,
\ee
which characterises the essential phase transition.
If we approach the BKT transition from the other side, {\it i.e.}\ 
within the massless phase, on large $L \times L$ lattices,
this picture predicts the magnetic susceptibility
$\chi$ to diverge as \cite{Kos74}
\be  \label{etar}
\chi
\propto L^{2 - \eta} \ (\ln L )^{-2r} \ , \quad
\eta_{\rm c} = 1/4 \ , \quad r_{\rm c} = -1/16 \ .
\ee
The numerical verification of the critical exponents 
$\eta_{\rm c}$ and $r_{\rm c}$ has been a long-standing challenge, 
despite the simplicity of the model, due to the tedious 
logarithmic finite size effects (see Ref.\ \cite{Kenna}
for an overview). The most satisfactory confirmation
of these values was reported in Ref.\ \cite{MHas}, based on 
simulations of the action (\ref{stanact}) on lattices
up to size $L = 2048$.\footnote{This study inserted the value 
$\eta_{\rm c} = 1/4$ as an input, and some ansatz  for the finite size 
scaling led to the thermodynamic extrapolation $r_{\rm c}= -0.056(7)$.}

The mapping of this system onto the sine-Gordon model also provides
analytic predictions for the Step Scaling Function \cite{Bal01}.
Again numerical simulations of the standard action yield a plausible
confirmation, if one refers to a specific ansatz for the finite size 
scaling \cite{Bal03}.

Lattice actions with additional spin couplings, such as
the Villain action \cite{Froh}, lead to the same continuum 
extrapolation, as expected due to the general principle
of {\em universality.} However, this property is less clear
for the highly unconventional {\em topological lattice actions}
\cite{topact}, which are invariant under (most) small deformations
of a spin configuration. The formulation of the 2d XY model by
topological lattice actions was recently discussed in Ref.\ 
\cite{XYtopact}. Here we address the most radical
variant, the {\em constraint action,} which does not have any
spin couplings at all. Instead the relative angles between
nearest neighbour spins are constrained to some maximum 
$\delta$.\footnote{The earlier history of $\sigma$ model 
simulations with a constraint angle includes Refs.\ 
\cite{prehist}.}
Hence the contribution of such a spin pair to the action amounts to
\be  \label{conact}
s (\vec e_{x}, \vec e_{y} ) = \left\{ \begin{array}{cccc}
0 &&& \vec e_{x} \cdot \vec e_{y} > \cos \delta \\
+ \infty &&& {\rm otherwise} \end{array} \right. \ .
\ee
Therefore all configurations which violate this constraint
for at least one spin pair $(\vec e_{x}$, $\vec e_{y})$ are excluded
from the functional integral, while all other configurations
have the same action $S [\vec e \, ] = \sum_{\langle x, y \rangle }
s (\vec e_{x}, \vec e_{y} ) = 0$. Obviously the concept of a classical
limit --- which corresponds to the action minimum ---
does not apply here, and perturbation theory neither. Instead
there is an enormous degeneracy, since all allowed configurations
have the same action.

For increasing $\delta$ a transition from a massless to a massive
phase was observed \cite{XYtopact}. In particular, at 
$\delta \gsim \delta_{\rm c}$ the correlation length could be 
fitted well to the behaviour analogous to relation (\ref{nue}),
\be  \label{deltadiv}
\xi \propto \exp \Big( \frac{\rm const.}
{\sqrt{\delta - \delta_{\rm c}}} \Big) \ .
\ee
This observation is based on measurements of $\xi$ on $L\times L$ 
lattices with $L=500$, $1000$ and $2000$. 
In the range $\delta = 1.89 \dots 2$, $\xi$ increases from 
$17.65(5)$ to $251.2(2)$. Referring to values which are converged
(for fixed $\delta$ and increasing $L$), we evaluated $\delta_{\rm c}$ 
by a fit to the ansatz (\ref{deltadiv}), which yields\footnote{In Ref.\
\cite{XYtopact} we gave the value $\delta_{\rm c} = 1.77521(57)$, 
but reconsidering the data and possible uncertainties, as well 
as fitting variants, we now conclude that the error might be larger. 
In the following sections we will present simulation results 
at the point, which would be critical in infinite volume; 
they refer to $\delta_{\rm c} = 1.77521$.\label{fndeltac}} 
\be  \label{deltac}
\delta_{\rm c} = 1.775(1) \ .
\ee
As another aspect, the evaluation of the critical exponent $\nu_{\rm c}$
will be discussed in Subsection 3.1.

Also the fits of the susceptibility $\chi$ to the form (\ref{etar})
attained a good quality for large volumes, $L = 128 \dots 4096$.
Hence we did observe a behaviour, which is compatible with a BKT
transition at $\delta_{\rm c}$, though this type of transition
could not be singled out unambiguously. The precision of that 
study was again limited by the logarithmic finite size effects. 

Here we are going to revisit this phase transition for the
constraint action. Section 2 deals with the helicity modulus.
Its numerical measurement is not straightforward for
topological lattice actions. It is achieved nevertheless 
by the use of dynamical boundary conditions. 
Section 3 addresses the second moment correlation length $\xi_{2}$ 
and the Binder cumulant $U_{4}$. We give results for the 
constraint action and for the standard action, which are 
compared to predictions of the spin wave approximation.
Section 4 discusses the statistics and correlations of vortices 
and anti-vortices, and the density of ``free vortices''. We 
study their dependence on the constraint angle $\delta$, and verify
the pair (un)binding mechanism for the BKT transition in
this unconventional formulation, where free vortices can 
appear without any energy cost (at $\delta > \pi/2$).
Three appendices are devoted to the algorithmic tools that we 
employed, and to the difficulty in formulating a cluster algorithm 
for simulations with dynamical boundary conditions.

\section{The helicity modulus}

The {\em helicity modulus} $\Upsilon$ is a quantity that
condensed matter physics often refers to. It is
sometimes also denoted as ``spin stiffness'' or ``spin rigidity'',
and it is proportional to the ``superfluid density''.
It appears in the literature on $O(4)$ models \cite{Fish} 
in a form, which would be called a Low Energy Constant
in the terminology of Chiral Perturbation Theory. In that
context, it is related to the pion decay constant $F_{\pi}$
as $\Upsilon \propto F_{\pi}^{2}$ \cite{HasLeu}.

$\Upsilon$ is a measure for the sensitivity of a system to torsion, 
{\it i.e.}\ to a {\em variation of a twist in the boundary conditions} 
\cite{Fish}. This is often a useful indicator to characterise a 
universality class, in addition to the critical exponents
and the Step Scaling Function.

\subsection{Definition and prediction}

For a conventional (non-topological) action on an 
$L \times L$ lattice, $\Upsilon$ can be defined as
\be  \label{standef}
\Upsilon 
= \frac{\partial^{2}}{\partial \alpha^{2}} 
F (\alpha ) |_{\alpha = 0} \ ,
\ee
where $F = - \frac{1}{\beta} \ln Z$ is the free energy, and
$Z$ is the partition function.
Here we assume the boundary conditions to be periodic in one 
direction, and twisted with the angle $\alpha$ in the other 
one. $F$ is minimal at $\alpha =0$, hence $\Upsilon$ corresponds 
to the curvature in this minimum.

Once we are dealing with topological lattice actions,
these notions need to be modified. In particular for the
constraint action (\ref{conact}) there is no coupling,
hence we consider the {\em dimensionless helicity modulus}
\be  \label{def}
\bar \Upsilon :=  \beta \, \Upsilon \ .
\ee
For the 2d XY model in a square volume, the critical value 
at the BKT transition was first predicted analytically \cite{Yc} 
as
$\bar \Upsilon_{\rm c,\, theory} = \frac{2}{\pi} . $
Later a tiny correction (below $0.2$ per mille) due to winding 
configurations was identified in Ref.\ \cite{Prok} (see also 
Ref.\ \cite{MHas}), which leads to
\be  \label{Upsctheo}
\bar \Upsilon_{\rm c,\, theory} = 
\frac{2}{\pi} \Big( 1 - 16 \pi e^{-4 \pi} \Big)
\simeq 0.636508 \ .
\ee

\subsection{Methods of the numerical measurement}

For the standard lattice action, there is a convenient way to 
measure $\bar \Upsilon$ at $\alpha =0$, such that the generation
of configurations can be restricted to periodic boundary conditions
\cite{Teit}.
The most extensive numerical study that evaluated $\bar \Upsilon$
in this way was performed by M.\ Hasenbusch \cite{MHas}; his results
are included in Figure \ref{figUpscrit}.
In his largest system, $L=2048$, he obtained at $\beta_{\rm c} = 
1.1199$ the value\footnote{By $\bar \Upsilon_{\rm c}$ we denote
the dimensionless helicity modulus at the critical parameter,
even in finite volume.}
$\bar \Upsilon_{\rm c} = 0.67246(10)$, which is still 
$5.6 ~\%$ too large. For the infinite volume extrapolation, he 
fitted his results for various sizes $L$ to the form
\be  \label{UpsLeq}
\bar \Upsilon_{\rm c} (L) = \bar \Upsilon_{\rm c,\, theory}
+ \frac{c_{1}}{\ln L + c_{2}} \ ,
\ee
with free parameters $c_{1}, \ c_{2}$, which worked decently.
Ref.\ \cite{MHas} also derived the theoretical prediction
\be  \label{c1Has}
c_{1} \simeq 0.3189
\ee 
in the spin wave limit, which was compatible with the fit 
to the data for the standard action. This 
prediction is based on the mapping onto the Gaussian model, where 
the parameter $\beta$ is inverted. Further arguments in favour of
the universality of the coefficient $c_{1}$ (though not of $c_{2}$) 
are based on the renormalisation group flow \cite{Peli}.\footnote{Moreover, 
even the sub-leading correction term $\propto \ln (\ln L ) / (\ln L)^{2}$ 
was worked out with renormalisation group techniques \cite{Peli}. 
However, we will see below that this term is not relevant for the 
discussion of our results with the constraint action, because they 
already deviate from the predicted leading order correction.}
Nevertheless it is interesting to reconsider the coefficient $c_{1}$ 
for a lattice action which does not involve any $\beta$ parameter.

For topological actions, the determination of $\bar \Upsilon$
at $\alpha =0$ fails.
A small change in $\alpha$ does (in general) not affect $F$ 
at all (in a finite volume). Naively referring to eq.\ (\ref{def})
would suggest $\bar \Upsilon = 0$.
Instead, a valid approach evaluates the curvature that eq.\ 
(\ref{def}) refers to from a {\em histogram} for the $\alpha$
values, which describes their probability $p( \alpha )$. 
Now $\alpha$ has to be treated as a {\em dynamical variable}
in the simulation. According to definitions (\ref{standef}),
(\ref{def}), its probability density is related to 
$\bar \Upsilon$ as \cite{OlsHol}
\be
\bar \Upsilon = - \frac{\partial^{2}}{\partial \alpha^{2}} \, 
\ln p (\alpha) \vert_{\alpha = 0} \ .
\ee
In practice the idea is to determine the curvature
in the maximum of $\ln p (\alpha)$ from a histogram up to
moderate $|\alpha | \, $. \\

If this model is formulated with the {\em step action} 
(which the literature calls ``step model'',
although the model is the same \cite{KenIrv,OlsHol}), the convenient
evaluation at $\alpha = 0$ is not applicable either.
Here the action of a pair of nearest neighbour spins is given by
\be
s (\phi_{x}, \phi_{y} ) = \left\{ \begin{array}{cccc}
- \beta &&& |\phi_{x} - \phi_{y} | < \pi / 2 \\
\ ~ \beta &&& {\rm otherwise} \end{array} \right. \ .
\ee
The BKT transition is observed around 
$\beta_{\rm c} \approx 1.2 \dots 1.3$ \cite{KenIrv,OlsHol}.
The corresponding histograms for $p(\alpha )$ have been studied 
in Ref.\ \cite{OlsHol}, which introduced twist angles in both directions.
In each direction it was divided into $L$ independent ``small twists'',
which were updated separately in a Metropolis simulation. 
In this way, Olsson and Holme measured, at $\beta_{\rm c}$, on a
$L=256$ lattice, $\bar \Upsilon_{\rm c} = 0.663(6)$. 
This is closer to the BKT value than the results for the
standard action (even in huge volumes \cite{MHas}),  
but still $4.2 \ \%$ too large. The authors of Ref.\ \cite{OlsHol}
were confident that a large $L$ extrapolation is compatible 
with the BKT prediction. They also fitted their data to the form 
(\ref{UpsLeq}) and inserted $c_{1}=1/\pi$ \cite{WebMin}, 
which is close to Hasenbusch's value 
(\ref{c1Has}).\footnote{Also this small correction is due to
configurations with non-zero winding number.}

In our case, we only deal with one twist at one of the boundaries,
an infinite step height, but a flexible step angle $\delta$.
We have to switch between spin updates and twist angle updates.
Since a Metropolis accept/reject step is fully deterministic,
we are guided to a {\em heat bath} algorithm, see Appendix A.
We update the spins or $\alpha$ one by one; the problem with 
the formulation of a {\em cluster algorithm} is discussed in
Appendix B.

\subsection{The helicity gap}

In infinite volume, the BKT theory predicts a discontinuity
of the helicity modulus. 
When we apply the 2d XY model to describe superfluids,
this jump has a direct physical interpretation:
$\bar \Upsilon$ is then related to the viscosity, which 
drops to $0$ in a discontinuous manner. 

\begin{figure}[h!]
\centering
\includegraphics[angle=0,width=.57\linewidth]{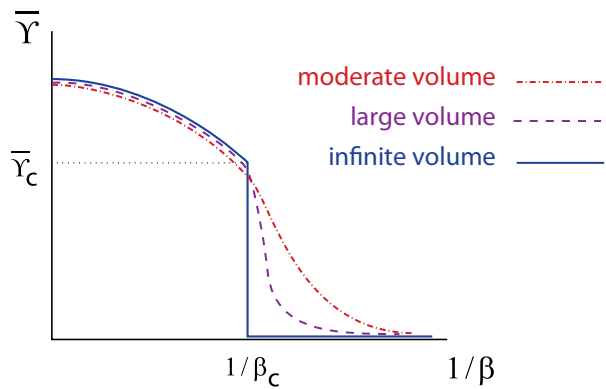}
\caption{\it A qualitative picture of the expected coupling
dependence of the helicity modulus in different volumes.}
\label{Upsquali}
\end{figure}

For lattice formulations with a coupling, this prediction implies 
that, as soon as the coupling exceeds its critical value, 
$\bar \Upsilon$ drops to $0$.
In finite volume the function $\bar \Upsilon (\beta )$
is continuous, but for increasing size $L$ the jump to $0$
is approximated better and better. This behaviour is
sketched qualitatively in Figure \ref{Upsquali}.
In fact, the observations for the standard action \cite{MinKim} 
and for the step action \cite{OlsHol} are compatible with this picture.

\begin{figure}[h!]
\centering
\includegraphics[angle=270,width=.85\linewidth]{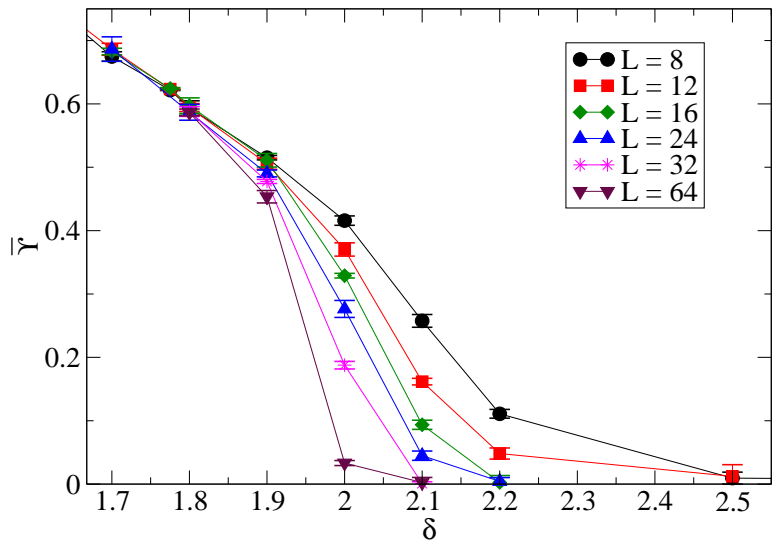}
\caption{\it Results for $\bar \Upsilon$ in six volumes, over a 
range of $\delta$ angles, which includes $\delta_{\rm c} \simeq 1.775$.
We observe the expected trend towards a jump down to $0$ next to 
$\delta_{\rm c}$, in analogy to the schematic Figure \ref{Upsquali}.}
\label{Upsdelta}
\end{figure}

\begin{figure}[h!]
\centering
\includegraphics[angle=0,width=.8\linewidth]{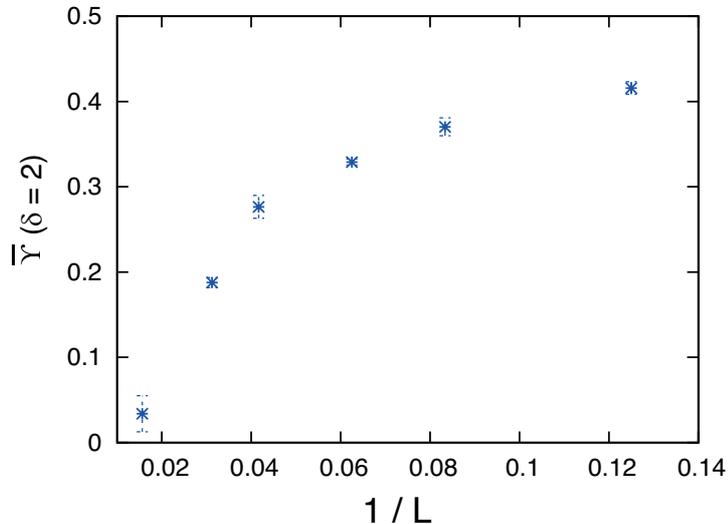}
\caption{\it Values of $\bar \Upsilon$ at $\delta = 2$ against $1/L$.
For increasing volume they rapidly drop towards $0$, in agreement
with the helicity gap picture.}
\label{Upsdelta2}
\end{figure}

We expect the same behaviour for the constraint action, 
where (in an infinite volume) $\bar \Upsilon (\delta )$ should 
jump to $0$ when $\delta$ exceeds $\delta_{\rm c}$. As a test, we measured
$\bar \Upsilon (\delta )$ in volumes of size $L= 8 \dots 64$. 
Figure \ref{Upsdelta} shows the results, which are well compatible 
with this expectation.
To further underscore this observation, Figure \ref{Upsdelta2}
shows specifically the values  $\bar \Upsilon (\delta =2)$
in various volumes, which have a highly plausible extrapolation
to $0$ in the thermodynamic limit $L\to \infty$.

\subsection{Critical value of the helicity modulus}

\begin{figure}[h!]
\centering
\includegraphics[angle=270,width=.8\linewidth]{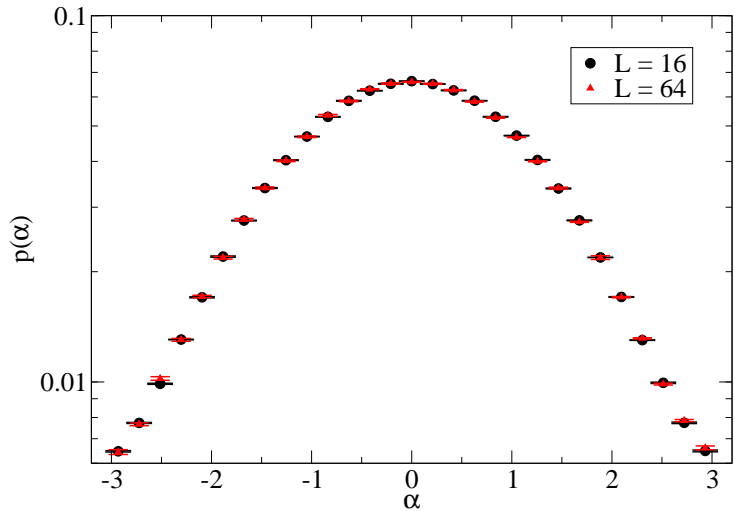}
\caption{\it Histograms for the twist angle $\alpha$,
measured by heat bath simulations in various volumes. A fit
in some interval $[-\alpha_{0}, \ \alpha_{0}]$ 
($\alpha_{0} = (n_{\rm bin}-1) \cdot \pi /30 $) determines 
the curvature in the maximum at $\alpha = 0$.}
\label{histo}
\end{figure}

We now focus on the critical constraint angle given in
Ref.\ \cite{XYtopact} (cf.\ Section 1), 
$\delta_{\rm c} = 1.77521$, and measure 
$\bar \Upsilon_{\rm c}$ in various volumes.
We simulated the model on $L \times L$ lattices in the range
$L = 8 \dots 256$ with dynamical boundary conditions.\footnote{If
we were at an exactly  massless point in any volume, the resulting 
curve $\bar \Upsilon_{\rm c}(L)$ should be universal, since $L$
is the only scale involved. However, since we fixed (as well as
possible) $\delta$ to
its value which is critical at $L \to \infty$, the correlation
length in finite volume will be finite, and we actually see
a combination of lattice artifacts and finite size effects.}
Figure \ref{histo} shows two of our histograms for the
twist angles $\alpha$. The histograms for $L \leq 128$ are based 
on more than a million $\alpha$ values, see Table \ref{tabUpscrit}.
The optimal evaluation of the curvature
at $\alpha =0$ was identified by probing a variety of bin sizes, 
and fitting ranges. The best option (regarding the ratio
$\chi^{2}$/d.o.f.) involves 31 bins for the $\alpha$ values.
A parabolic fit is performed by including the 
number $n_{\rm bin}$ of bins around $\alpha =0$, which again
minimises the quantity $\chi^{2}$/d.o.f.
Table \ref{tabUpscrit} displays this number for each volume,
along with our results for $\bar \Upsilon_{\rm c}$ and 
their uncertainties.
\begin{table}[h!]
\renewcommand{\arraystretch}{1.2}
\centering
\begin{tabular}{|c||r|c|l||c|c|}
\hline
$L$ & statistics & $n_{\rm bin}$ & $\chi^{2}$/d.o.f. 
& $\bar \Upsilon_{\rm c}$ & error \\
\hline
\hline
  8  & 16\,097\,744 & 21 & 0.000004 & 0.6217  &  0.0006 \\
\hline
 12  &  9\,142\,032 & 17 & 0.000004 & 0.6234  &  0.0011 \\
\hline
 16  &  6\,266\,902 & 15 & 0.000005 & 0.6243  &  0.0015 \\
\hline
 32  &  2\,506\,795 & 13 & 0.000006 & 0.6257  &  0.0025 \\
\hline
 64  &  1\,001\,355 & 13 & 0.000016 & 0.6355  &  0.0041 \\
\hline
128  &  1\,2417\,90 &  7 & 0.000010 & 0.6345  &  0.0041 \\
\hline
256  &   584\,178 &  7 & 0.000020 & 0.6333  &  0.0073 \\
\hline
\end{tabular}
\caption{\it Numerical results for the dimensionless helicity 
modulus, measured at the critical constraint angle $\delta_{\rm c}$,
in $L \times L$ volumes, by means of histograms for the twist angles
$\alpha$ (the statistics gives the number of $\alpha$ values involved).}
\label{tabUpscrit}
\end{table}

\begin{figure}[h!]
\includegraphics[angle=270,width=1.\linewidth]{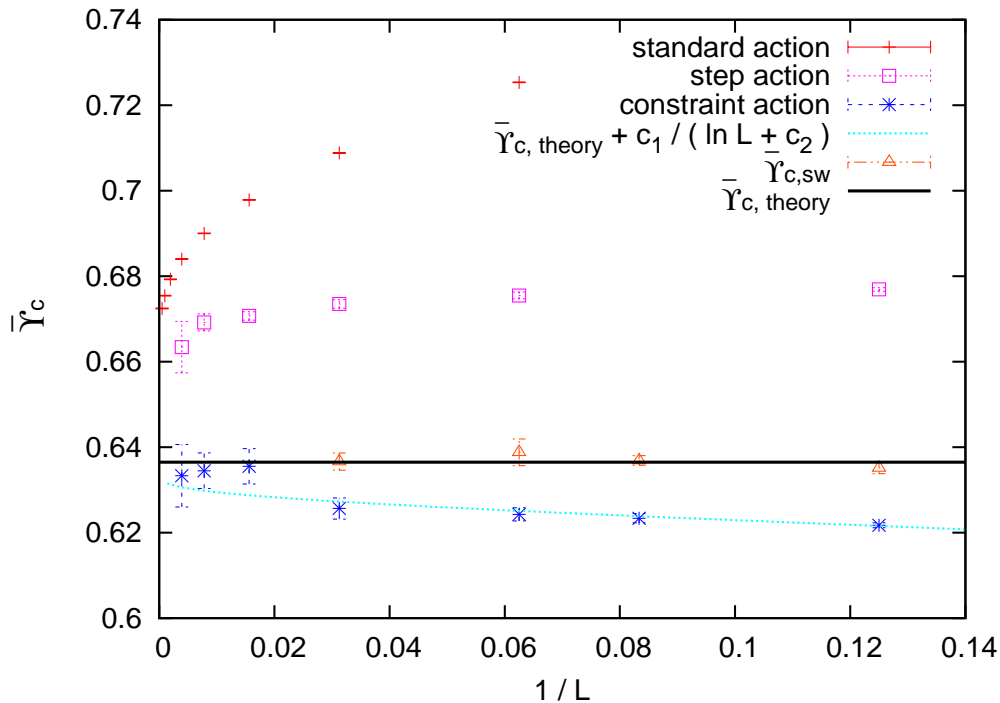}
\caption{\it Numerical results for $\bar \Upsilon_{\rm c}$ on $L \times L$
lattices at the critical parameter, for the standard 
action (data from Ref.\ \cite{MHas}), the step action (data from 
Ref.\ \cite{OlsHol}) and for the constraint action (data from
Table \ref{tabUpscrit}). In the former two cases, all measured 
results differ significantly from the theoretical BKT value 
$\bar \Upsilon_{\rm c,\, theory}$ in eq.\ (\ref{Upsctheo}), which could 
only be attained with extended extrapolations. Only for the constraint 
action the results agree with the BKT prediction for $L \geq 64$, and 
the deviation is just $1.9 \ \%$ ($2.3 \ \%$) even at $L=16$ ($L=8$). 
For that action, we show the fit to eq.\ (\ref{UpsLeq}) 
with $c_{1} = -0.0335382$, $c_{2}= 0.166499$. (We also include
data measured at $\delta_{\rm c,sw}=1.756$, denoted as 
$\bar \Upsilon_{\rm c,sw}$, which will be discussed in Subsection 3.1.)}
\label{figUpscrit}
\end{figure}

For all sizes $L$ that we consider, the deviations of
$\bar \Upsilon_{\rm c}$ from the theoretical value 
$\bar \Upsilon_{\rm c,\, theory }$ in eq.\ (\ref{Upsctheo}) 
is less than $ 2.4 \ \%$, and for $L \geq 64$
our results confirm the prediction within the errors.
This observation is highly remarkable in view of earlier attempts
to measure $\bar \Upsilon_{\rm c}$ with other lattice actions,
which could only claim agreement with the BTK value based on 
specific 
large volume extrapolations.  For illustration,
Figure \ref{figUpscrit} compares our results to those for the 
standard action \cite{MHas} and for the step action \cite{OlsHol}.

The fit for the constraint action data to eq.\ (\ref{UpsLeq}) yields
\be  \label{c1c2crit}
c_{1} = -0.034(11) \ , \quad c_{2} = 0.17(80) \ .
\ee
The corresponding graph is included in Figure \ref{figUpscrit}.

Thus there is a clear discrepancy from the $c_{1}$ value in eq.\ 
(\ref{c1Has}), which was derived in the spin wave limit and with
renormalisation group considerations.
Of course, there is no rigorous guarantee that our data
really reveal the asymptotic large $L$ behaviour, although
this is what one would naturally expect.

Still, this discrepancy may look surprising, and it calls for
a discussion. On the other hand, this observation
might be viewed in light of the recent
experience with topological (and mixed) lattice actions:
\begin{itemize}

\item Based on the mapping of the 2d XY model onto the sine-Gordon
model, Ref.\ \cite{Balog01} derived --- in addition to
the continuum Step Scaling Function \cite{Bal01} ---
the coefficient of the leading lattice artifact term, 
which was also assumed to be universal. In fact, it agrees with 
data for the standard action \cite{Bal03}, but it clearly disagrees
with results for various topological actions, including the
constraint action \cite{XYtopact}.

\item In the large $N$ limit of the 2d O(N) model, 
there are corrections to the continuum Step Scaling Functions,
which start in the quadratic order of the lattice spacing,
multiplied by some logarithmic factor. In this case, even
the leading power of this logarithmic term differs between the 
standard action and topological actions \cite{drastic}.

\item For the 2d O(3) model with a ``mixed action'' (with a standard
coupling plus an angular constraint), the Step Scaling Function
has lattice artifacts, which seem incompatible with the expected
asymptotic agreement with the standard action artifacts \cite{drastic}.

\item For the 1d O(2) model (the quantum rotor), the topological
actions are plagued by linear lattice artifacts \cite{topact}, 
although one might expect them to be generically quadratic. 
(However, since this is not a field theoretic example, universality 
arguments do not apply.)

\end{itemize}

Nevertheless, in the next section we are going to investigate further
the applicability of spin wave predictions to the constraint
action results, now proceeding to much larger lattices.

\section{Binder cumulant and second moment correlation length}

For the measurement of the dimensionless helicity modulus $\bar \Upsilon$ 
we were restricted to use the heat bath algorithm described in 
Appendix A. Therefore the lattice sizes which could be reached were 
rather moderate ($L \leq 256$). In this range, the results agree
with the thermodynamic prediction for $\bar \Upsilon_{\rm c,theory}$, 
but not with the prediction for the coefficient $c_{1}$.

For an extended test of the applicability of spin wave theory
predictions, we now consider the {\em Binder cumulant} 
$U_{4}$  and the {\em second moment correlation length} $\xi_{2}$. 
This can be done at fixed periodic boundary conditions,
hence we can apply the more efficient Wolff cluster algorithm 
\cite{Wolff} and explore much larger lattices.

We start from the magnetisation $\vec{m}$
and the magnetic susceptibility  $\chi$,
\begin{equation} 
\vec{m} =  \sum_x \vec{e}_x  \quad , \quad
\chi =  \frac{1}{L^2} \langle \vec{m}^2 \rangle \ .
\label{chi}
\end{equation}
The Binder cumulant $U_4$ is obtained as\footnote{Here we follow
the notation of Ref.\ \cite{Has08}, which differs from the
wide-spread convention $U_{4} = 1 - \frac{1}{3}
\langle (\vec{m}^{2})^{2} \rangle / \langle \vec{m}^{2} \rangle^{2}$.}
\begin{equation}
 U_{4} = \frac{\langle (\vec{m}^{2})^{2} \rangle}
{\langle \vec{m}^{2} \rangle^{2}} \ .
\end{equation}

To compute the second moment correlation length $\xi_{2}$, one 
considers the Fourier transform of the correlation function 
$G(x-y) = \langle \vec e_x \cdot \vec e_y \rangle$
\begin{equation}
\label{ftrans}
\widetilde G(p) = \sum_x G(x) \exp({\rm i} p x) \ ,
\end{equation}
on an $L \times L$ lattice. It contains the
magnetic susceptibility $\chi = \widetilde G(p=0)$, as well as the 
quantity $F = \widetilde G(p=(2\pi /L, 0))$ at the 
smallest non-vanishing momentum. The second moment 
correlation length is given by
\begin{equation}
\label{second}
\xi_{2}=\frac{1}{2 \sin(\pi/L)} \left(\frac{\chi}{F}-1\right)^{1/2} \ .
\end{equation}
It is very similar to the (actual) correlation length, but
easier to measure.

Predictions for $U_4$ \cite{Has08} and $\xi_{2}/L$ \cite{MHas}
in the thermodynamic limit $L \to \infty$ of the 2d XY model
have been calculated by using the spin wave approximation,
which yields
\bea
U_{4, {\rm sw}} &=& 1.018192(6) + \frac{C_{1}}{\ln L + C_{2}} 
+ \dots \ , \nn \\
\xi_{2, {\rm sw}} / L &=& 0.7506912 \dots + \frac{C_{1}{'}}
{\ln L +C_{2}{'}} + \dots \ , \label{U4xi2}
\eea
where $C_{1}, \, C_{2}, \, C_{1}{'}, C_{2}{'}$ are constants. 
These asymptotic values, and the coefficients $C_{1}$, $C_{1}{'}$
are interesting in view of universality.
Ref.\ \cite{Has08} obtained results for the standard action of 
the 2d XY model, which are consistent with two other models in 
the same universality class, over a wide range of sizes $L$.

\begin{figure}[h!]
\begin{center}
\includegraphics[width=0.6\textwidth,angle=270]{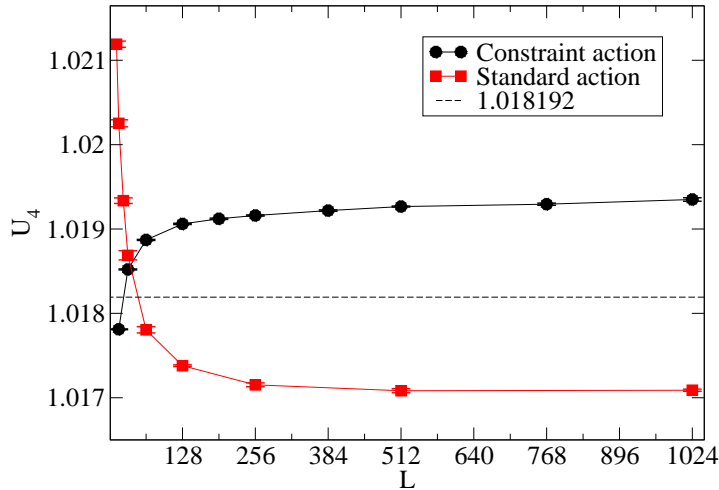}
\vspace*{-2mm}
\caption{\it The Binder cumulant for the constraint action 
and for the standard action, compared to the spin wave prediction
in the large volume limit.}
\label{binderboth}
\end{center}
\vspace*{-4mm}
\end{figure}

We computed $U_4 $ and $\xi_{2}/L$ for the constraint action, as 
well as the standard action, at the critical parameter, on $L \times L$ 
lattices in the range $L = 12 \dots 1024$. Our results 
for the standard action (at $\beta_{\rm c} = 1.1199$) fully 
agree with those obtained by Hasenbusch \cite{Has08}. 
Figure \ref{binderboth} shows our results for the Binder cumulant.
For both lattice actions we observe a trend to a plateau value, 
which is close to the prediction (\ref{U4xi2}), but not in exact 
agreement. There remains a deviation of about 1 per mille, which
is positive (negative) for the constraint action (standard action).

\begin{figure}[h!]
\begin{center}
\includegraphics[width=0.6\textwidth,angle=270]{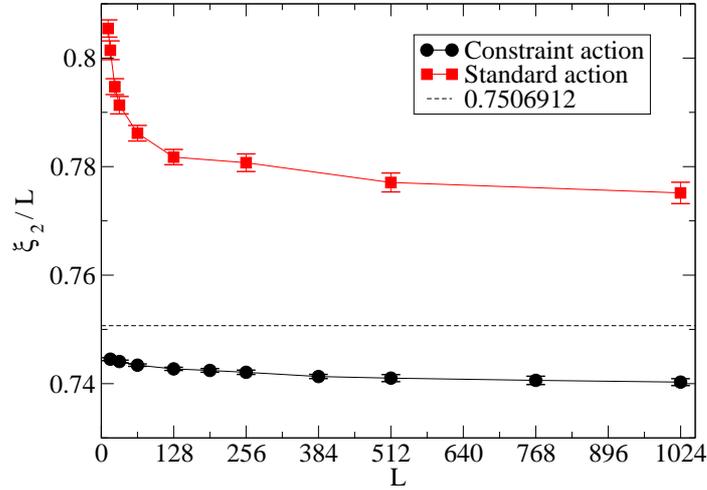}
\vspace*{-2mm}
\caption{\it The second moment correlation length $\xi_{2}$, divided by 
the lattice size $L$, for the standard action and for the constraint 
action. The results for both actions deviate a little from the
spin wave prediction.}
\label{xi2both}
\end{center}
\end{figure}

The behaviour of the ratio $\xi_{2}/L$ is qualitatively similar,
as Figure \ref{xi2both} shows. Here we observe a deviation from
the spin wave prediction (\ref{U4xi2}) on the percent level, again
with opposite signs for the two actions. This time the data for 
the constraint action are clearly closer to the prediction.
Our numerical values for $U_{4}$ and $\xi_{2}/L$  are listed 
in Tables \ref{numconstraint} and \ref{numstandard}.

\begin{table}[h!]
\renewcommand{\arraystretch}{1.2}
\centering
\begin{tabular}{|c||c|c|}
\hline
$L$ &  $U_4$ & $\xi_{2}/L$ \\
\hline
\hline
  16  & 1.017811(5) & 0.7445(3)  \\
\hline
  32  & 1.018521(6)  &  0.7441(3)  \\
\hline
  64  & 1.018871(7)  & 0.7434(3)  \\
\hline
  128  & 1.019061(8)  & 0.7427(4) \\
\hline
  192  & 1.019121(9)  & 0.7424(4)  \\
 \hline
  256  & 1.01916(7)  & 0.7421(4)  \\ 
\hline
  384  & 1.019218(9)  & 0.7413(4)    \\  
\hline
  512  & 1.019266(6)  & 0.7410(7)   \\
\hline
  768  & 1.019294(13)  & 0.7406(8)    \\  
\hline
  1024  & 1.01935(21)  & 0.7402(7)   \\
\hline
\end{tabular}
\caption{\it Numerical results for $U_{4}$ and $\xi_{2}/L$,
obtained with the constraint action.}
\label{numconstraint}
\end{table}

\begin{table}[h!]
\renewcommand{\arraystretch}{1.2}
\centering
\begin{tabular}{|c||c|c|}
\hline
$L$ &  $U_4$ & $\xi_{2}/L$ \\
\hline
\hline
  12  & 1.02119(4) & 0.8054(17)  \\
\hline
  16  & 1.02025(5) & 0.8014(18)  \\
\hline
  24  & 1.01933(4) & 0.7947(15)  \\
\hline
  32  & 1.01868(6) & 0.7913(18) \\
\hline
  64  & 1.01780(4) & 0.7861(15)  \\
\hline
  128  & 1.01738(2) & 0.7817(15) \\
\hline
  256  & 1.01715(3) & 0.7807(17)  \\
\hline
  512  & 1.01708(3) & 0.7771(18) \\
\hline
  1024  & 1.01708(2) & 0.7751(21) \\
\hline
\end{tabular}
\caption{\it Numerical results for $U_{4}$ and $\xi_{2}/L$,
obtained with the standard action.}
\label{numstandard}
\end{table}

Considering the stable behaviour on the largest lattice sizes
that we explored, and the even larger lattices where 
Hasenbusch simulated the standard action \cite{Has08}, it 
is not obvious to expect 
that both curves would ultimately converge 
to the predicted values for $L\to \infty$, even in the presence of 
logarithmic finite size effects. Hasenbusch suspected that (universal)
sub-leading finite $L$ corrections to the spin wave results could explain 
this discrepancy. 
However, this scenario has now the additional difficulty to explain
our observation that the results for different lattice actions 
disagree as well.\footnote{What is expected to differ are the
parameters $C_{2}$, $C_{2}{'}$, as well as higher order terms,
so the question is if this can explain the deviating results. 
In the range up to $L=1024$ this seems unlikely.}

For the standard action, Hasenbusch estimated a minimum of $U_{4}$ 
around $L \approx 6200$, and an extremely slow increase on even 
larger lattices. Of course, this scenario --- and its analogue 
for the constraint action with a $U_{4}$ maximum in some huge
volume --- cannot be excluded. 
The alternative scenario would be that 
$U_{4}$ and $\xi_{2}/L$ are not truly 
(but still approximately) universal quantities.
It remains as an open question if that alternative scenario 
holds, and how it could possibly be explained.

\subsection{Criticality determined from spin wave predictions}

For lattice sizes in the range up to $L=1024$, we have seen in 
Figures \ref{binderboth} and \ref{xi2both}, as well as Table 
\ref{numconstraint} and \ref{numstandard}, small but significant 
discrepancies from the spin wave predictions for the quantities 
$U_{4}$ and $\xi_{2}/L$. In this subsection we are going to
explore an alternative approach, which takes these predictions
as a basis to determine the critical point. Hence in this alternative
approach we assume the predictions (\ref{U4xi2}) to be correct,
and the numerical data to display a visible convergence towards 
these values for the lattice sizes under consideration.

\begin{figure}[h!]
\centering
\includegraphics[angle=270,width=.7\linewidth]{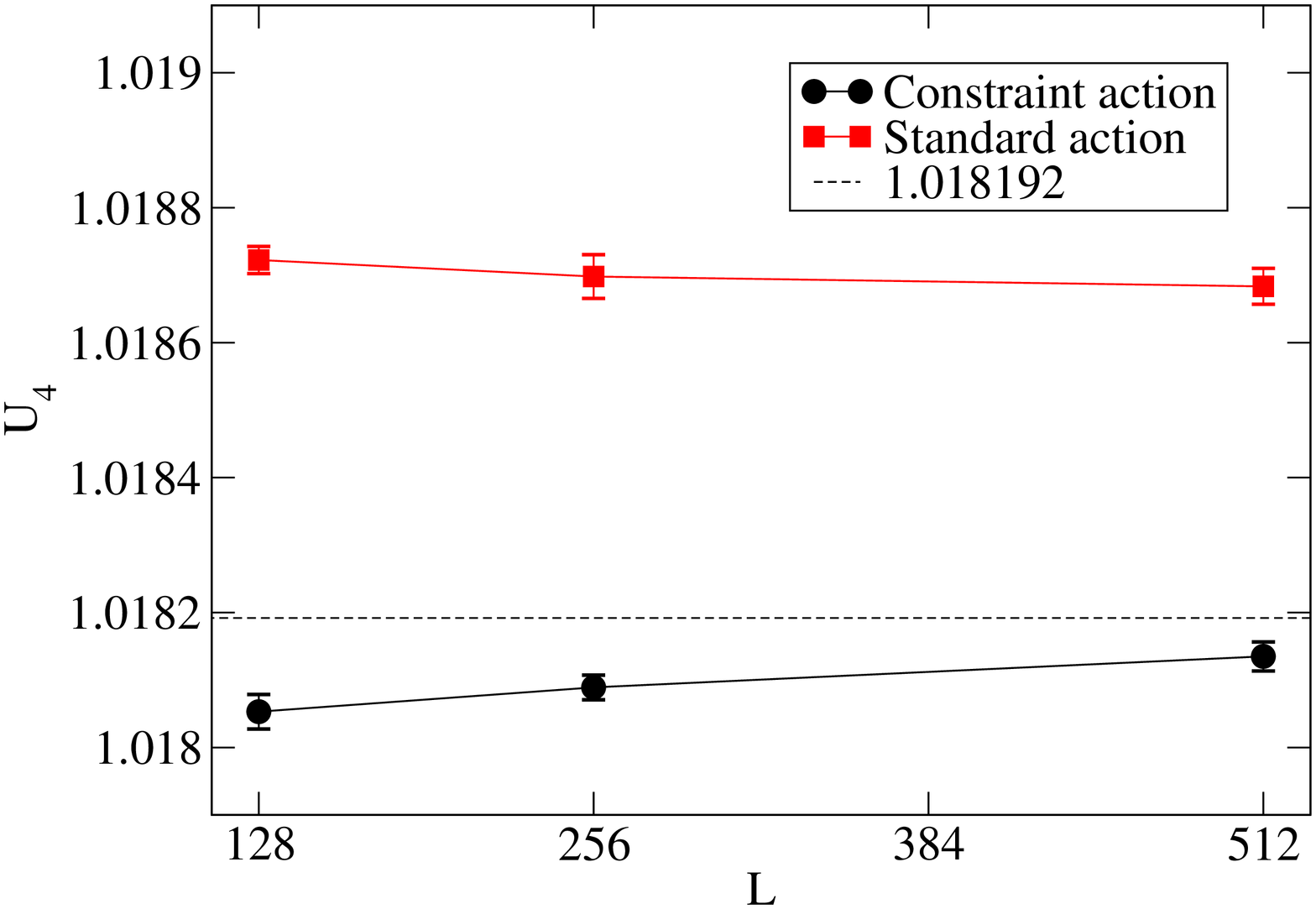}
\vspace*{-5mm} \\
\includegraphics[angle=270,width=.7\linewidth]{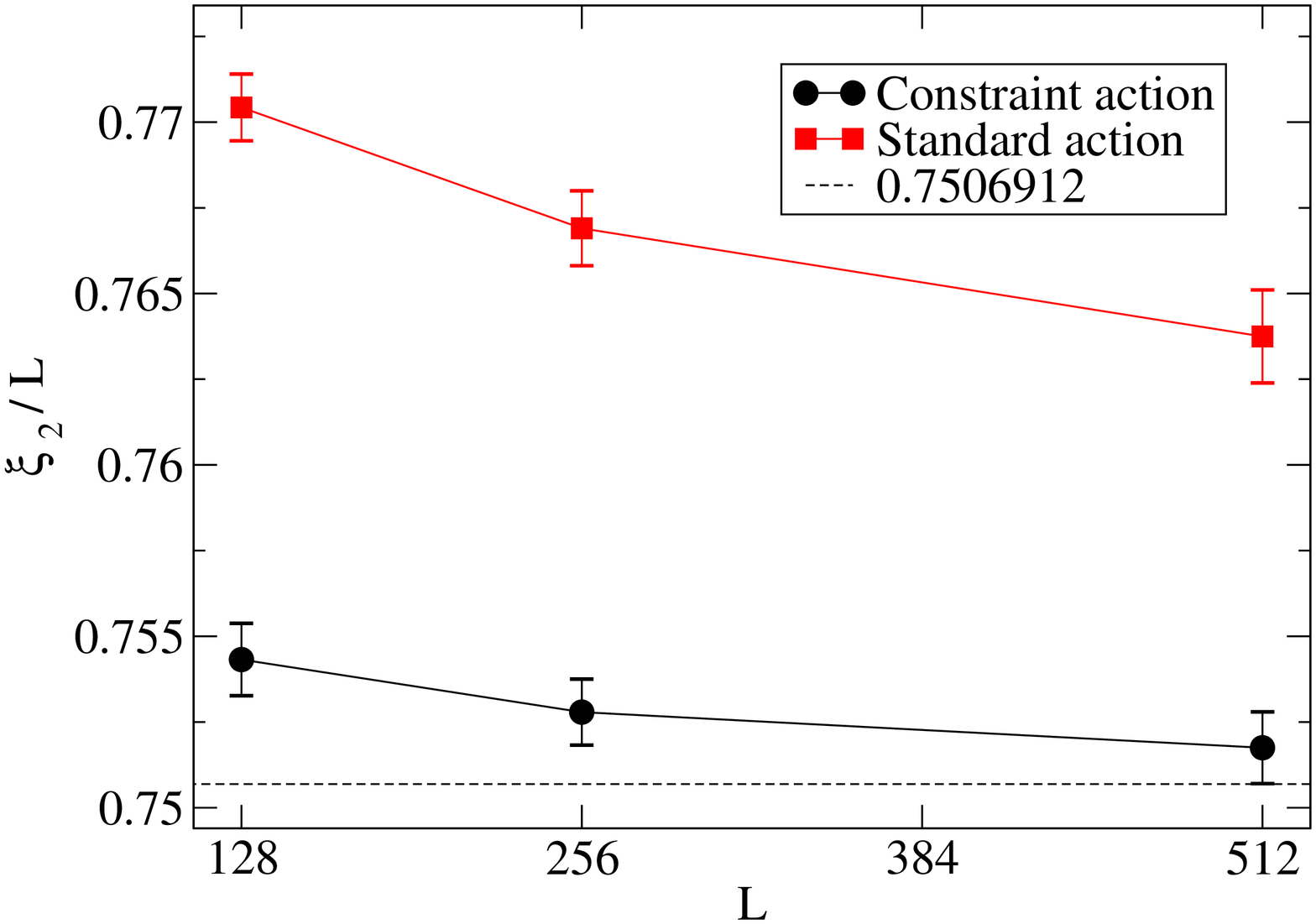}
\caption{\it The Binder cumulant $U_{4}$ and the
ratio $\xi_{2}/L$ at the alternative suggestions
for the critical parameters, based on the spin wave predictions:
$\delta_{\rm c,sw} = 1.756$ (for the constraint action) and
$\beta_{\rm c,sw} = 1.112$ (for the standard action).}
\label{swcritfigs}
\end{figure}

For the constraint action this leads to an alternative
suggestion for the critical angle, which amounts to
\be
\delta_{\rm c,sw} = 1.756 (2) \ . 
\ee
The results for $U_{4}$ and $\xi_{2}/L$ at this angle are shown 
in Figure \ref{swcritfigs}. Indeed this shift from $\delta_{\rm c}$ to
$\delta_{\rm c,sw}$ provides convincing convergence of {\em both}
quantities towards the spin wave predictions.

Next we measured $\bar \Upsilon$ at $\delta_{\rm c,sw}$ to verify
if the curve $\bar \Upsilon_{\rm c,sw} (L)$ also converges to
the predicted value $\bar \Upsilon_{\rm c,theory}$ 
in the thermodynamics limit. 
These data points are included in Figure \ref{figUpscrit}. They
do converge to $\bar \Upsilon_{\rm c,theory}$, which is already attained
(within errors) at $L=12$. So to this point $\delta_{\rm c,sw}$ looks
like a plausible alternative for the critical constraint angle.

However, if we fit these data to the predicted asymptotic formula
(\ref{UpsLeq}) we obtain
\be \label{c1c2sw}
c_{1} = 0.00015(12) \ , \quad c_{2} = -2.19(9) \ ,
\ee
so the discrepancy from the $c_{1}$ prediction (\ref{c1Has})
persists.\footnote{On the other hand, $c_{2}$ differs strongly
from its value in eq.\ (\ref{c1c2crit}), but (unlike $c_{1}$)
that parameter is not claimed to be universal.}
In fact the determination of $\delta_{\rm c}$ from $\xi$ is somewhat
involved, but even if we shift this value considerably, the 
absolute value $|c_{1}|$ --- as obtained from our fits ---
remains tiny and incompatible with $\approx 0.32$. Of course
we cannot rigorously rule out that the use of tremendously large 
volumes and ultra-high precision would still provide that value.
However, the results accessible with a reasonable 
computational effort can hardly be reconciled with the $c_{1}$ 
prediction, even with some tolerance for the critical $\delta$ angle.

In order to explore the alternative approach of this subsection
further, we tried the same method also for the standard action.
However, in this case it is not possible to find any $\beta$
parameter which would lead to good convergence of $U_{4}$ {\em and}
of $\xi_{2}/L$ towards the the values in eq.\ (\ref{U4xi2}). As
a compromise we choose $\beta_{\rm c,sw} = 1.112$; a separate
matching of $U_{4}$ (of $\xi_{2}/L$) would suggest a larger (lower)
$\beta$, as Figure \ref{swcritfigs} shows. Moreover, $\beta_{\rm c,sw}$
is incompatible with the value that we quoted earlier in
Section 1, $\beta_{\rm c} = 1.1199(1)$ \cite{HasPin}, which is based
on a careful high precision study, and which is accepted in the
literature. 

\begin{figure}[h!]
\centering
\includegraphics[angle=270,width=.7\linewidth]{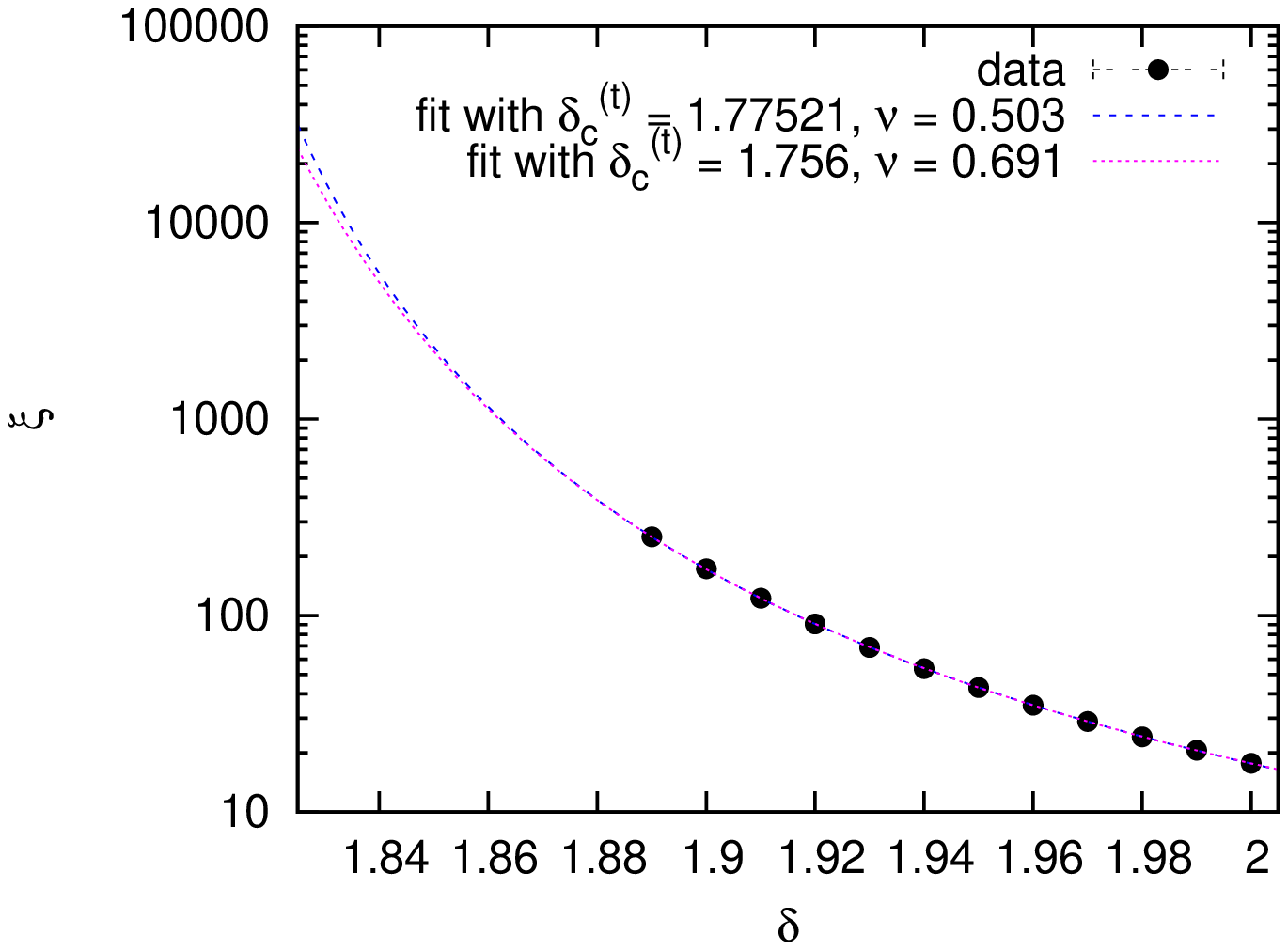}
\includegraphics[angle=270,width=.7\linewidth]{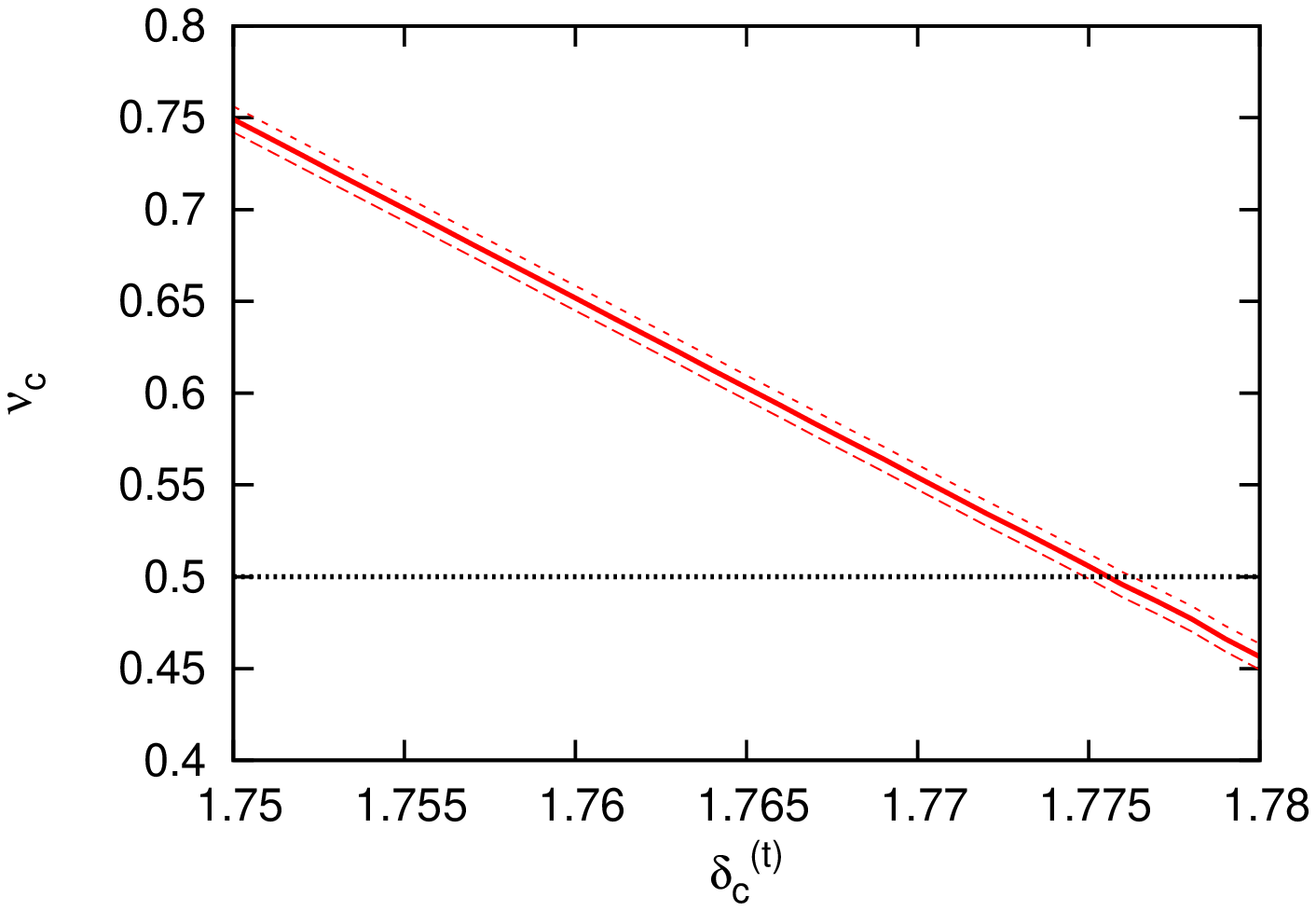}
\caption{\it On top: 
The correlation length $\xi$, measured in large volumes
(up to convergence in $L$ at fixed $\delta$) and its fits to the
function (\ref{critexpnu}) for $\delta_{\rm c}^{\rm (t)} = \delta_{\rm c}
= 1.77521$ and for $\delta_{\rm c}^{\rm (t)} = \delta_{\rm c,sw} = 1.756$.
Below: The critical exponent $\nu_{\rm c}$, obtained by fits
to eq.\ (\ref{critexpnu}), as a function of the angle 
inserted for $\delta_{\rm c}^{\rm (t)}$.
We see that $\delta_{\rm c}^{(t)} \in [1.7748 , 1.7763]$ 
leads to agreement with the theoretically predicted exponent 
$\nu_{\rm c} = 1/2$ \cite{Kos74}, in contrast to $\delta_{\rm c,sw}$.}
\label{nucritfigs}
\end{figure}

Therefore we consider the critical parameter evaluated
from the correlation length $\xi$ more reliable.
As a further cross-check, we fit the formula
\be  \label{critexpnu}
\xi \propto \exp \Big( {\rm const.} /
(\delta - \delta_{\rm c}^{\rm (t)})^{\nu_{\rm c}} \Big) 
\ee
to our data for the correlation length $\xi$ at $\delta = 1.89 \dots 2$.
Now we insert some trial value for $\delta_{\rm c}^{(t)}$ and evaluate the 
critical exponent $\nu_{\rm c}$ through the fit. In particular, if we 
insert $\delta_{\rm c}^{\rm (t)} = \delta_{\rm c} = 1.77521$ or
$\delta_{\rm c}^{\rm (t)} = \delta_{\rm c,sw} = 1.756$, 
we obtain $\nu_{\rm c} = 0.501(7)$ and $0.691(7)$, respectively.
The corresponding extrapolations in $\xi (\delta )$ are shown
in Figure \ref{nucritfigs} (above). The plot below illustrates
the results for $\nu_{\rm c}$ obtained in this way
(along with its error band) over the entire range 
$\delta_{\rm c}^{\rm (t)} = 1.75 \dots 1.78$. The BKT value $\nu_{\rm c}=1/2$
\cite{Kos74} singles out $1.7748 < \delta_{\rm c} < 1.7763$,
which is clearly incompatible with $\delta_{\rm c,sw}$.

\section{Free vortices and vortex--anti-vortex pairs}

We define the relative angle between nearest neighbour spins
with a ${\rm mod}~2\pi$ operation, which acts such that
the absolute value becomes minimal,
\be  \label{DeltaPhi}
\Delta \Phi_{x,y} = (\Phi_{x} - \Phi_{y}) \ {\rm mod} \ 2 \pi
\in ( - \pi , \pi ] \ .
\ee
If we sum these relative angles over the corners $x,y,z,w$ of 
a plaquette, and normalise by $2\pi$, we obtain the {\em vortex 
number}
\be
v_{\Box} = \frac{1}{2 \pi}
( \Delta_{x,y} + \Delta_{y,z} + \Delta_{z,w} + \Delta_{w,x})
\in \{ 1, 0, -1 \} \ .
\ee
For $v_{\Box}=1$ ($v_{\Box}=-1$) the plaquette carries a vortex
(an anti-vortex); higher vortex numbers ($|v_{\Box}| >1$) 
do not occur.

As in the previous section we deal with periodic boundary 
conditions. Hence Stokes' Theorem implies that the total 
vorticity always vanishes,
\be
\sum_{\Box} v_{\Box} = 0 \ .
\ee

With this terminology, the picture of vortex (un)binding
as the mechanism that drives the BKT transition can be probed
explicitly. For the standard action (\ref{stanact}) this has been
done in Refs.\ \cite{Jap,ToboChe,BoHuJa,GupBai}, and the results
are essentially consistent with the suggested picture.
These considerations were static, comparing the behaviour
in the different phases. 

There are also studies of the dynamics of the 
unbinding when $\beta$ decreases gradually below 
$\beta_{\rm c}$. Such considerations proceeded first by solving 
the Fokker-Planck equation \cite{ChuWil}, and later
by Monte Carlo simulations \cite{JelCug}. For the
standard action, the outcome was again compatible with the
picture of dissociating vortex--anti-vortex pairs. \\

\begin{figure}[h!]
\centering
\includegraphics[angle=270,width=0.8\linewidth]{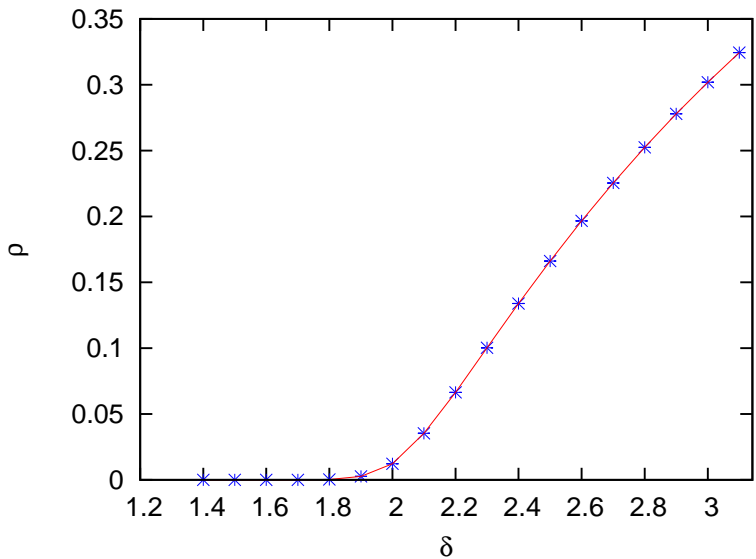}
\caption{\it The vortex density $\rho$, {\it i.e.}\ the number 
of plaquettes with $|v_{\Box}| = 1$, divided by the volume. We show 
results obtained at $L=128$, but the volume dependence is marginal. 
Error bars are included, but too small to be seen, and 
the line is drawn to guide the eye.}
\label{Vdensity}
\end{figure}

For the constraint action that we are investigating here,
we first show in Table \ref{Vdensetab} and Figure \ref{Vdensity} 
how the total vortex plus anti-vortex density $\rho$ 
depends on $\delta$.\footnote{For the corresponding density
with the standard action we refer to Ref.\ \cite{Ota2}.} 
Vortices are possible for $\delta > \pi / 2$, 
but their density become significant only around 
$\delta \gsim 1.9$, {\it i.e.}\ somewhat above $\delta_{\rm c}$.

\begin{figure}[h!]
\centering
\includegraphics[angle=0,width=0.46\linewidth]{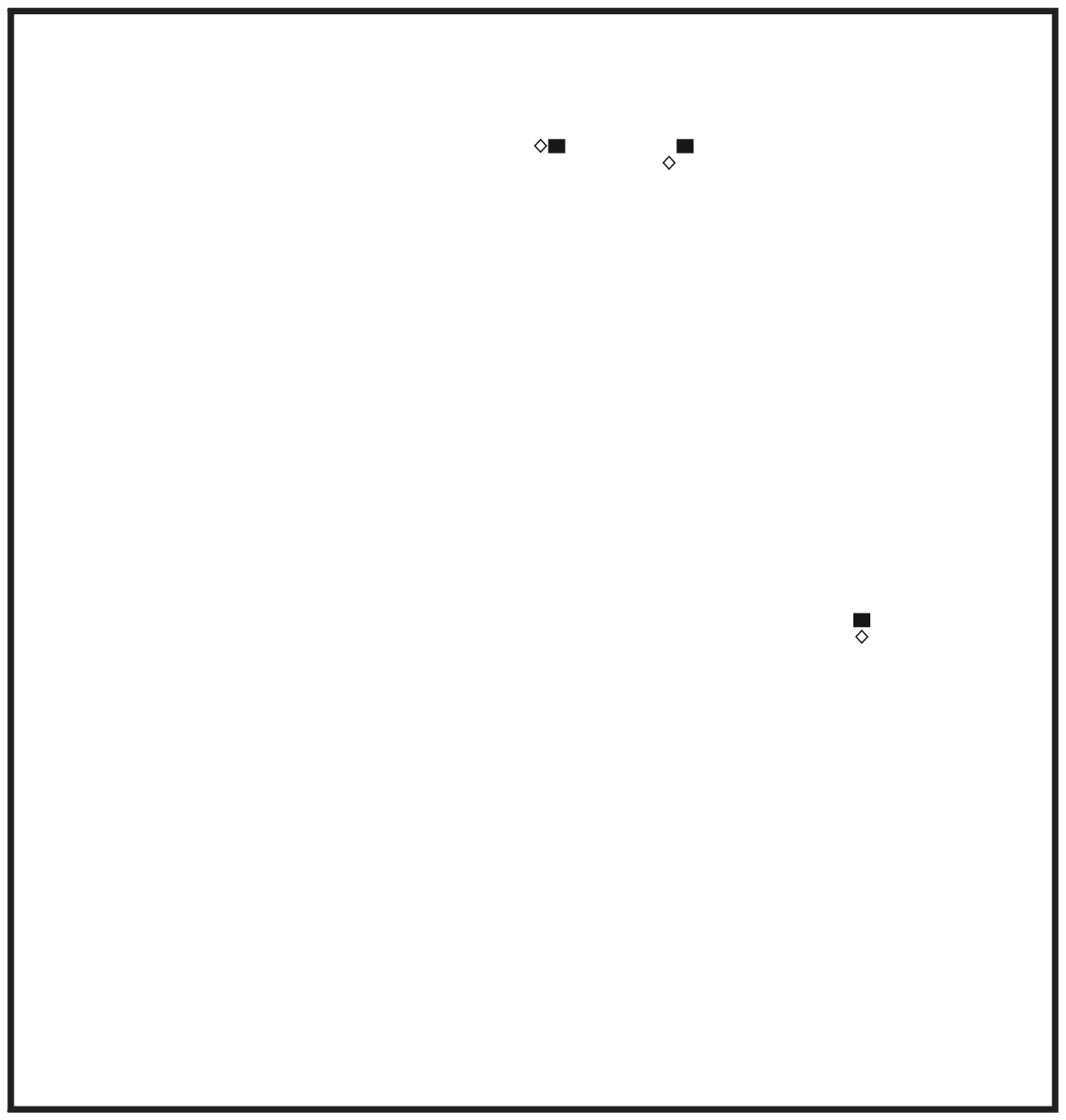}
\hspace*{2mm}
\includegraphics[angle=0,width=0.46\linewidth]{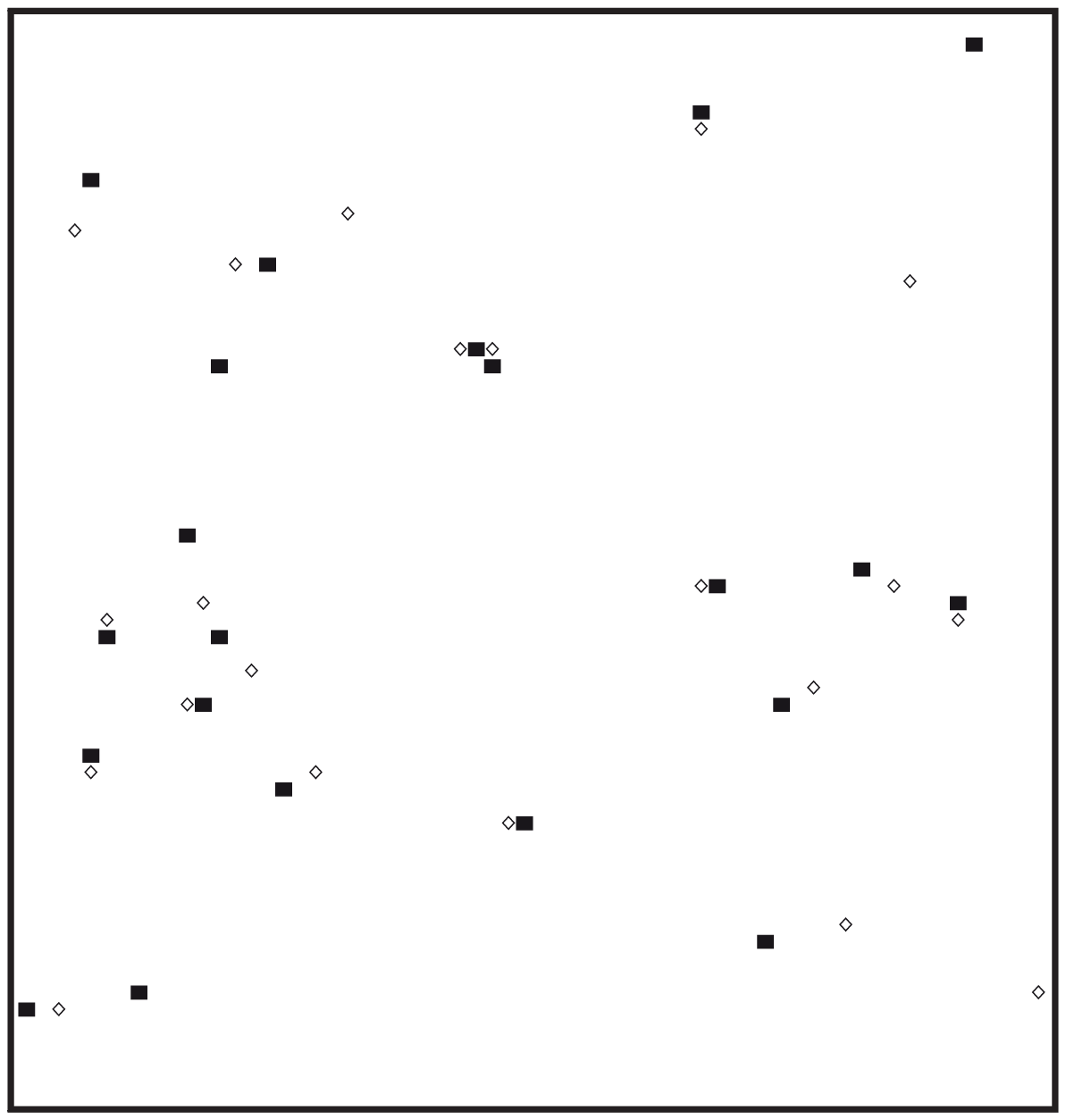}
\vspace*{2mm} \\
\includegraphics[angle=0,width=0.46\linewidth]{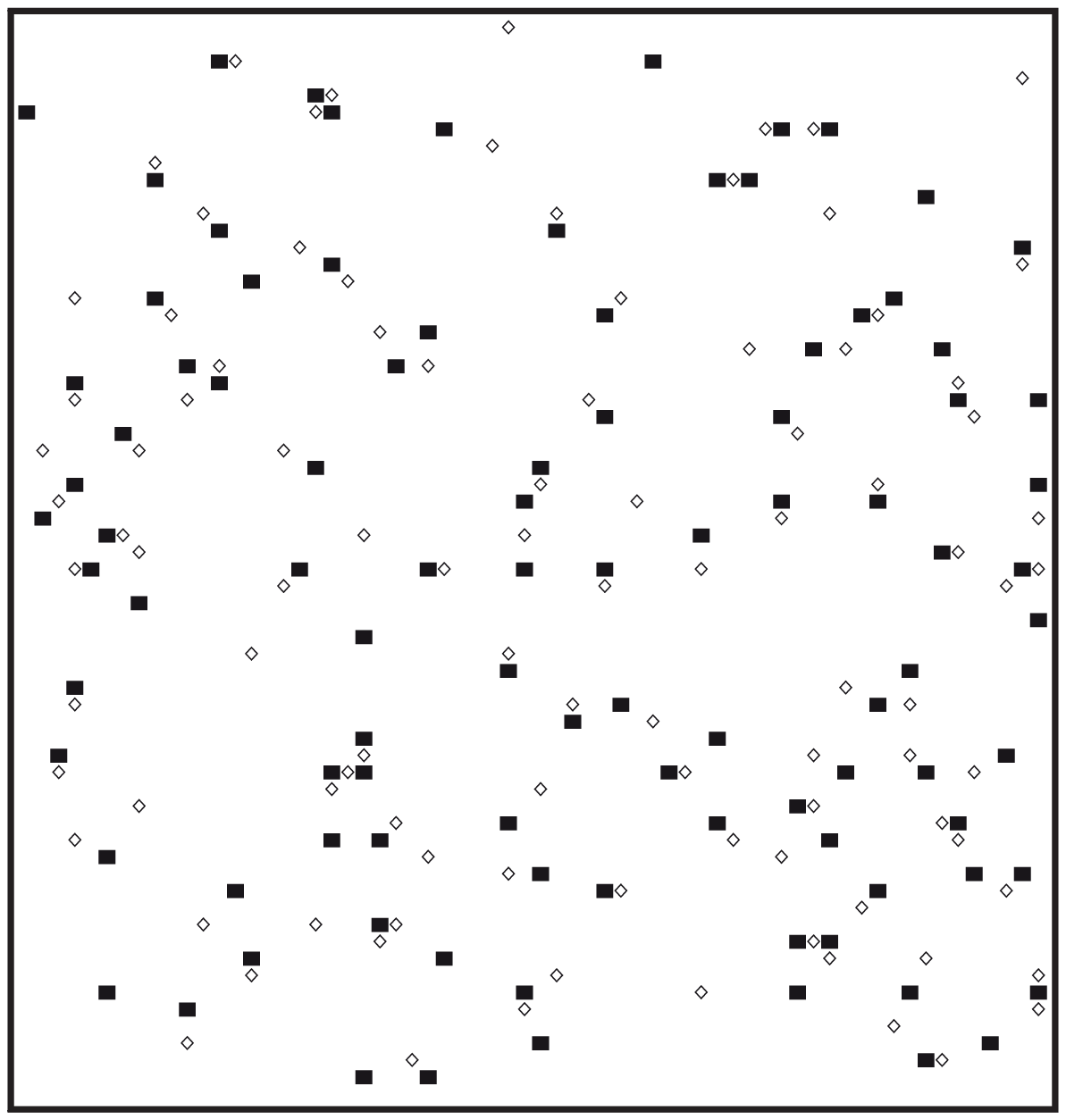}
\hspace*{2mm}
\includegraphics[angle=0,width=0.46\linewidth]{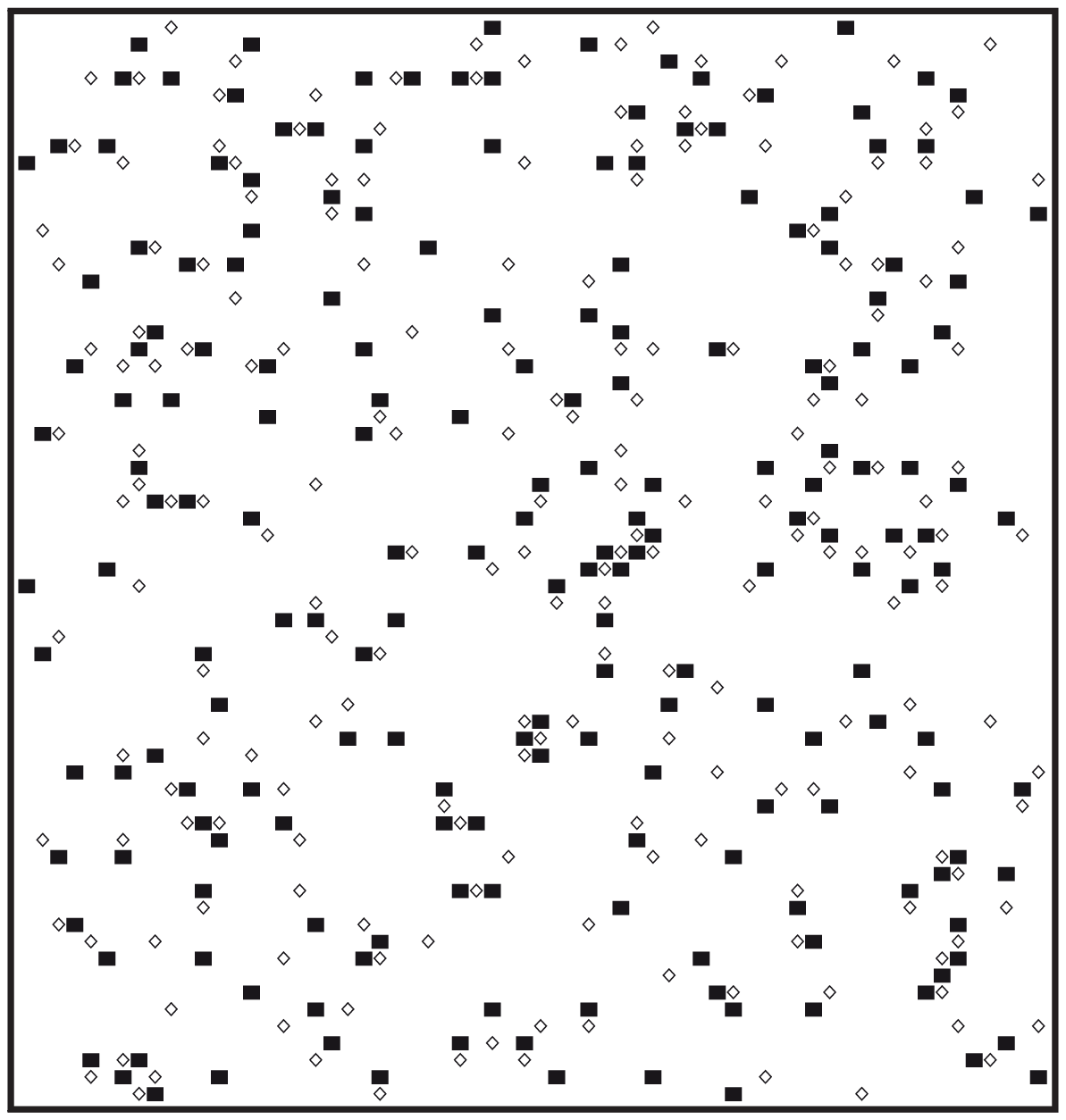}
\caption{\it The distribution of vortices (filled squares) 
and anti-vortices (empty diamonds) in typical configurations 
on a $64 \times 64$ lattice at $\delta =1.85$, $2$ (above), 
and $\delta = 2.15$, $2.3$ (below). For growing $\delta$ we observe 
an increasing vortex density (cf.\ Figure \ref{Vdensity}), but a 
decreasing trend towards vortex--anti-vortex pair formations, and 
therefore more and more free vortices (cf.\ Figure \ref{freeVdensity}).}
\label{Vmap}
\end{figure}

\begin{table}[h!]
\renewcommand{\arraystretch}{1.1}
\centering
\begin{tabular}{|c||c||c|c|c|}
\hline
$\delta$ & $\rho$ & $\rho_{1}^{\rm free}$ & 
$\rho_{2}^{\rm free}$ & $\rho_{4}^{\rm free}$ \\ 
\hline
\hline
1.8 & 0.00032(1)  & 0.00016(5) & 0.000021(7) & 0.000013(5) \\
\hline
2   & 0.0122(1)   & 0.0068(1)  & 0.0045(2)   & 0.0016(1) \\
\hline
2.2 & 0.066424(7) & 0.0334(1)  & 0.015517(8) & 0.00172(1) \\
\hline
2.4 & 0.13390(5)  & 0.05527(3) & 0.01733(3)  & 0.00030(1) \\
\hline
2.6 & 0.19661(2)  & 0.06701(2) & 0.014110(3) & 0.000032(1) \\
\hline
\end{tabular}
\caption{\it The density of all vortices (plus anti-vortices), $\rho$, 
and of the ``free vortices'' $\rho_{r}^{\rm free}$ --- without any 
opposite partner within Euclidean distance $r = 1$, $2$ or $4$ --- 
at different constraint angles $\delta$. 
The measurement was performed at $L=128$, but the size dependence 
is modest.}
\label{Vdensetab}
\end{table}
As an illustration, we show in Figure \ref{Vmap} the vortex 
and anti-vortex distribution in typical configurations of a
$L=64$ lattice at $\delta =1.85$, $2$, $2.15$ and $2.3$.
We observe also here the increase in the total vortex density.
In addition we recognise a strong trend towards
vortex--anti-vortex pair formations at $\delta =1.85$,
which fades away for increasing $\delta$.
These specific configurations appear qualitatively
consistent with the (un)binding mechanism. However, a solid 
verification requires statistical investigations, which we will
present in the rest of this section.

\subsection{Free vortex density}

\begin{figure}[h!]
\centering
\includegraphics[angle=270,width=0.75\linewidth]{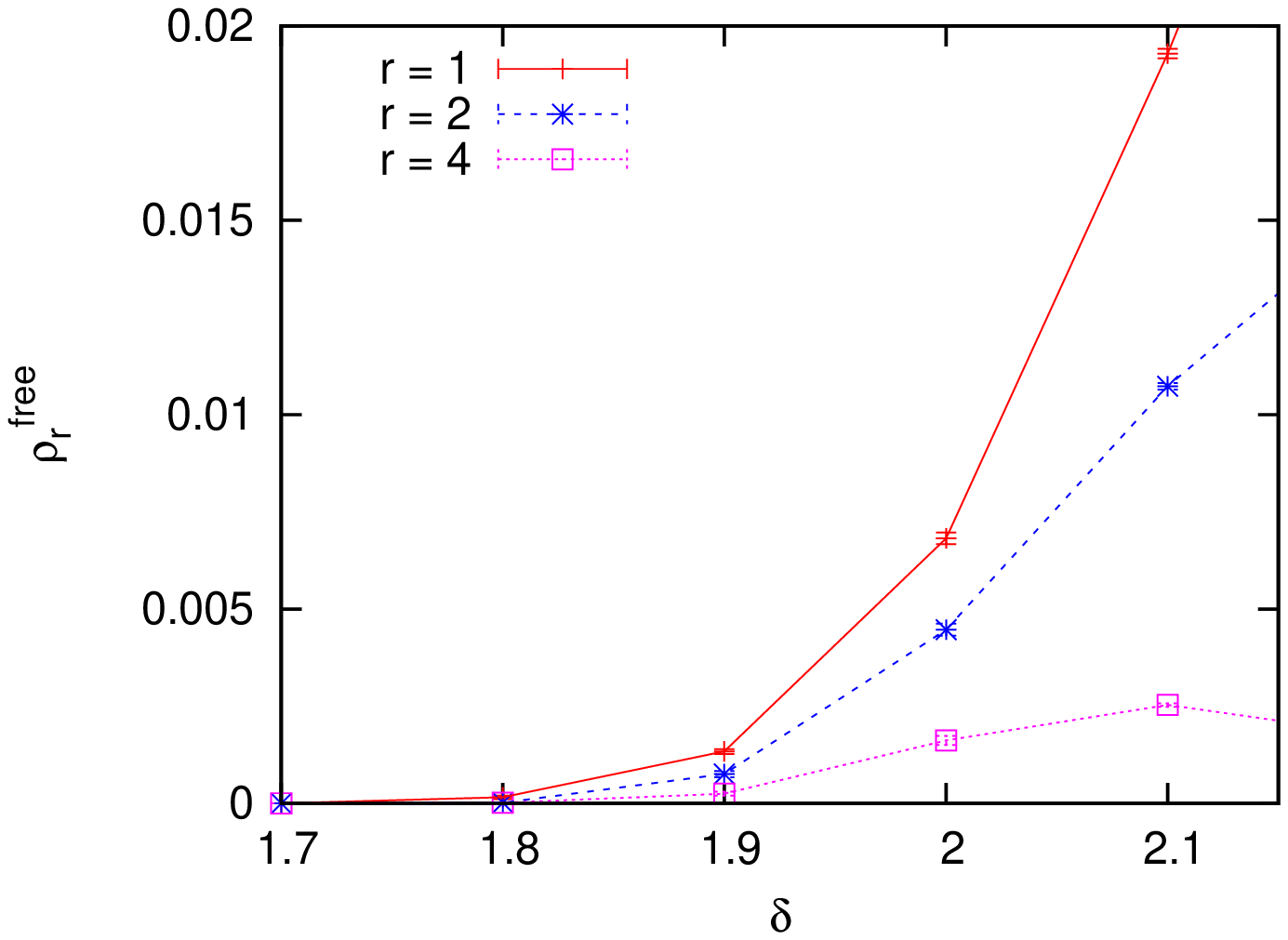}
\includegraphics[angle=270,width=0.75\linewidth]{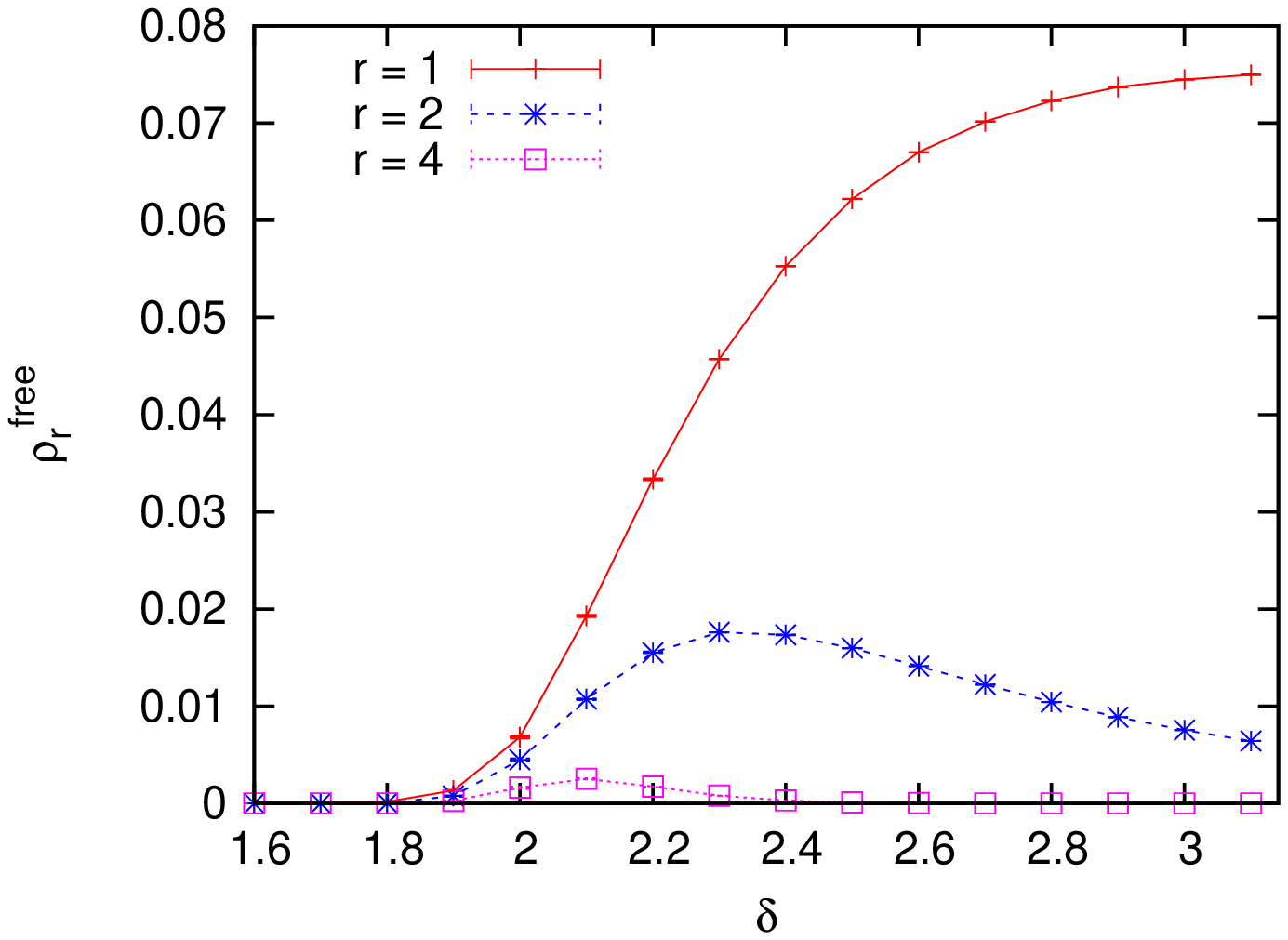}
\caption{\it The density of ``free vortices'', $\rho_{r}^{\rm free}$, 
{\it i.e.}\ vortices or anti-vortices, which do not have an opposite 
partner within some Euclidean distance $r$. 
For various choices of $r$, a significant density sets in
around $\delta = 1.8 \dots 1.9$. This is best seen from the plot above,
while the plot below provides an overview.
These results are obtained at $L=128$, but they are hardly size 
dependent.}
\label{freeVdensity}
\end{figure}

According to the established picture, it is not the
total vortex density which matters for
the fate of a long-ranged order, but rather the density of ``free 
vortices''. There is clearly some ambiguity in an explicit 
definition of this term. An obvious possibility is to count 
those vortices which are not accompanied by any anti-vortex 
(or vice versa) within some Euclidean distance $r$.
Table \ref{Vdensetab} and Figure \ref{freeVdensity}
show the corresponding free vortex densities 
$\rho_{r}^{\rm free}$ for $r=1$, $r=2$ and $r=4$.
In all cases, there is a significant onset in the interval $\delta 
= 1.8 \dots 1.9$, which is compatible with the onset of
$\rho$. For $r \geq 2$ the densities $\rho_{r}^{\rm free}$ 
decrease again at large $\delta$ angles, 
as a consequence of the high total density $\rho$. 
However, it is the first onset which indicates
the dissociation of vortex--anti-vortex pairs, and which is
therefore relevant for the BKT picture. Our observation
is compatible with this picture, up to a shift of the onset 
somewhat into the massive phase. A similar behaviour has been 
observed for the standard action \cite{Jap,ToboChe,BoHuJa,GupBai}.

\subsection{Vorticity correlation}

\begin{figure}[h!]
\centering
\includegraphics[angle=270,width=0.85\linewidth]{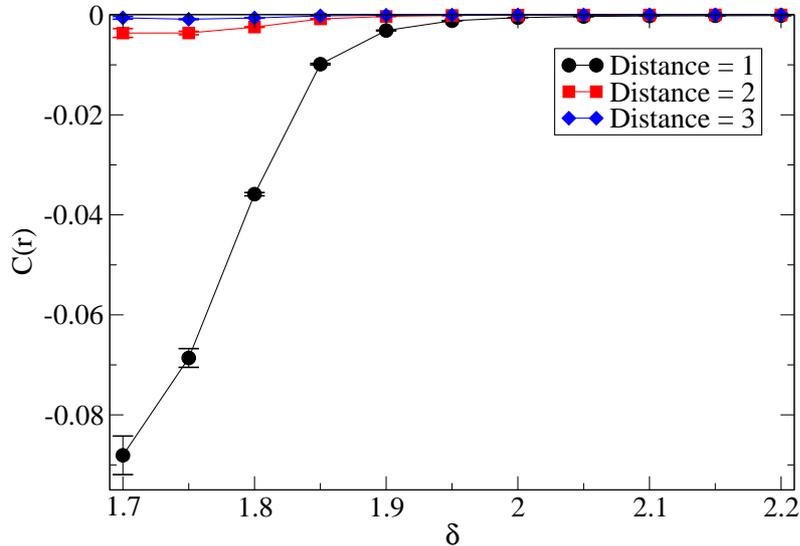}
\caption{\it The vorticity correlation function $C(r)$ of eq.\ 
(\ref{vorcor}) over distances $r=1,2,3$, as a function of the 
constraint angle $\delta$.
In particular, there is a strong anti-correlation over the minimal
distance $r=1$ at $\delta \lsim \delta_{\rm c}$, in agreement 
with the formation of vortex--anti-vortex pairs. 
This effect get lost rapidly as $\delta$ increases, since
the pairs tend to dissociate.}
\label{vorcorr}
\end{figure}

The established picture of the BKT transition also implies a sizable
vorticity anti-correlation over short distances in the massless
phase --- in particular over distance $1$.
Figure \ref{vorcorr} shows results for the vorticity 
correlation function\footnote{A similar consideration with
the standard action is given in Ref.\ \cite{Ota2}.}
\be  \label{vorcor}
C(r) = \langle \, v_{\Box , (x_{1},x_{2})} \, v_{\Box , (x_{1}+r,x_{2})} \, 
\rangle |_{|v_{\Box , (x_{1},x_{2})}|=1}
\ee
over distances $r = 1$, 2 and 3, at a set of constraint angles
$\delta \geq \delta_{\rm c}$. Indeed we confirm a marked 
anti-correlation over distance 1 around $\delta_{\rm c}$,
which rapidly fades away when $\delta$ increases.

Of course, there is a trivial argument for $C(1) < 0$,
since the relative angle of the common link of adjacent 
plaquettes contributes to their vorticity with opposite sign.
However, this alone does not explain the sharp
slope at $\delta = 1.8 \dots 1.9$, and the remaining 
anti-correlation at $r=2$ and $3$.

This observation is reminiscent of the anti-correlation of 
the topological charge density in gauge theories,
for configurations in the neutral sector \cite{fixtopo2}. 
This property was confirmed explicitly on the lattice for 
QCD \cite{chitQCDNf2} and for the Schwinger model \cite{BHSV}.

\subsection{Vortex--anti-vortex pair formation}

Finally we want to investigate the vortex--anti-vortex
pair formation from yet another, more direct perspective.
Given a configuration, we first identify its $N$ vortices 
and $N$ anti-vortices, and we search for the {\em optimal 
pairing.} This optimisation minimises the quantity
\be \label{D2}
D^{2} = \frac{1}{N} \, \sum_{i=1}^{N} d_{{\rm VA},\, i}^{\, 2} \ \, ,
\ee
where $d_{{\rm VA},\, i}$ are the Euclidean distances that separate
the vortex--anti-vortex partners. The straightforward method
of checking all possibilities is safe, but only applicable
up to $N \approx 14$. We work again at lattice size
$L = 128$ with constraint angles up to $\delta =2.05$, where
typically $N$ is close to $200$ (see Figure \ref{Vdensity}). 
In order to still identify the optimal
pairing (with high probability), we applied the 
technique of ``simulated annealing'', see Appendix C.

\begin{figure}[h!]
\vspace*{-4mm}
\centering
\includegraphics[angle=270,width=0.85\linewidth]{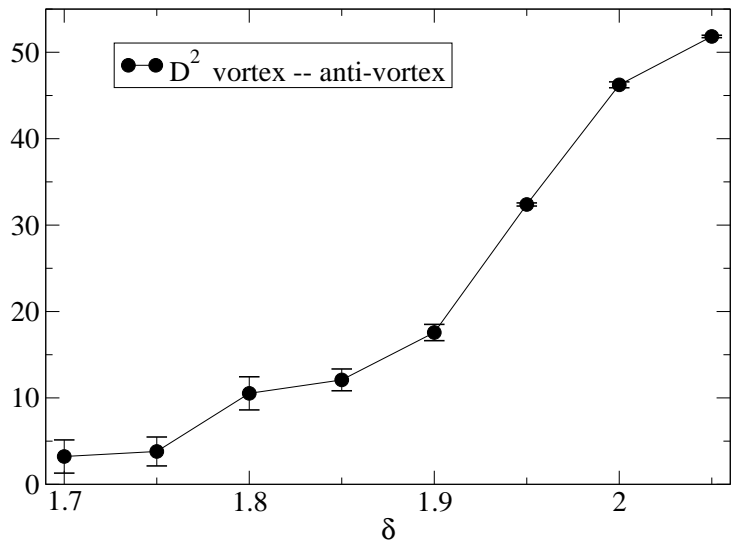}
\caption{\it The mean distance squared within the vortex--anti-vortex 
pairs, $D^{2}$, on a $L=128$ lattice. The significant (drastic) increase 
above $\delta_{\rm c}$ (above $\delta =1.9$) indicates the unbinding 
of some (numerous) pairs.}
\label{annealplot}
\end{figure}

The results for $D^{2}$ at different
angles $\delta$ are shown in Figure \ref{annealplot}.
We see a slight increase at $\delta \gsim 1.8$,
followed by a sharp increase at $\delta \gsim 1.9$. This is
another piece of evidence for the (un)binding mechanism
behind the BTK transition, and once more the obvious effect 
is shifted somewhat into the massless phase.

\begin{figure}[h!]
\centering
\vspace*{-5mm}
\includegraphics[angle=270,width=0.7\linewidth]{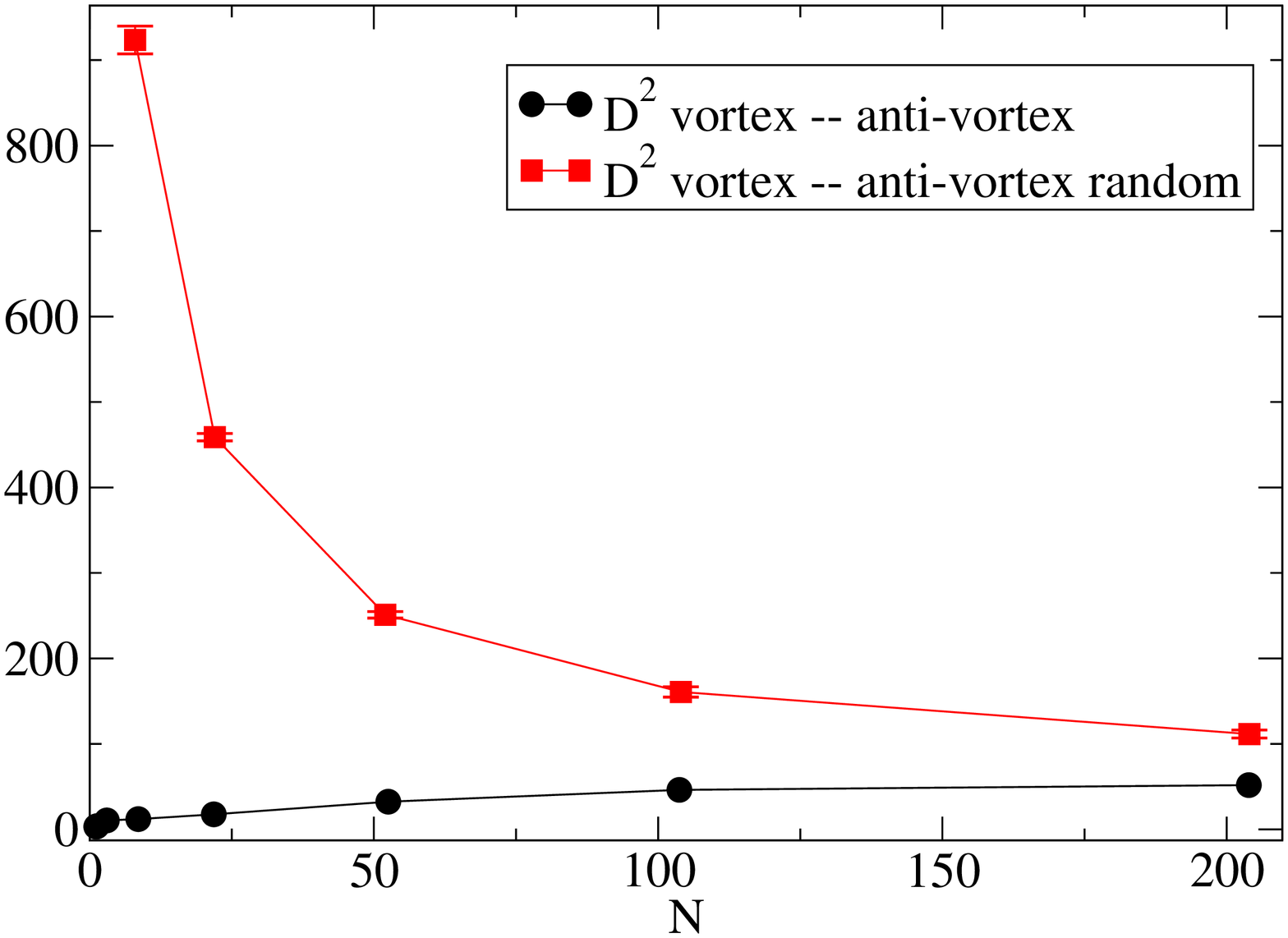}
\vspace*{-5mm} \\
\includegraphics[angle=270,width=0.7\linewidth]{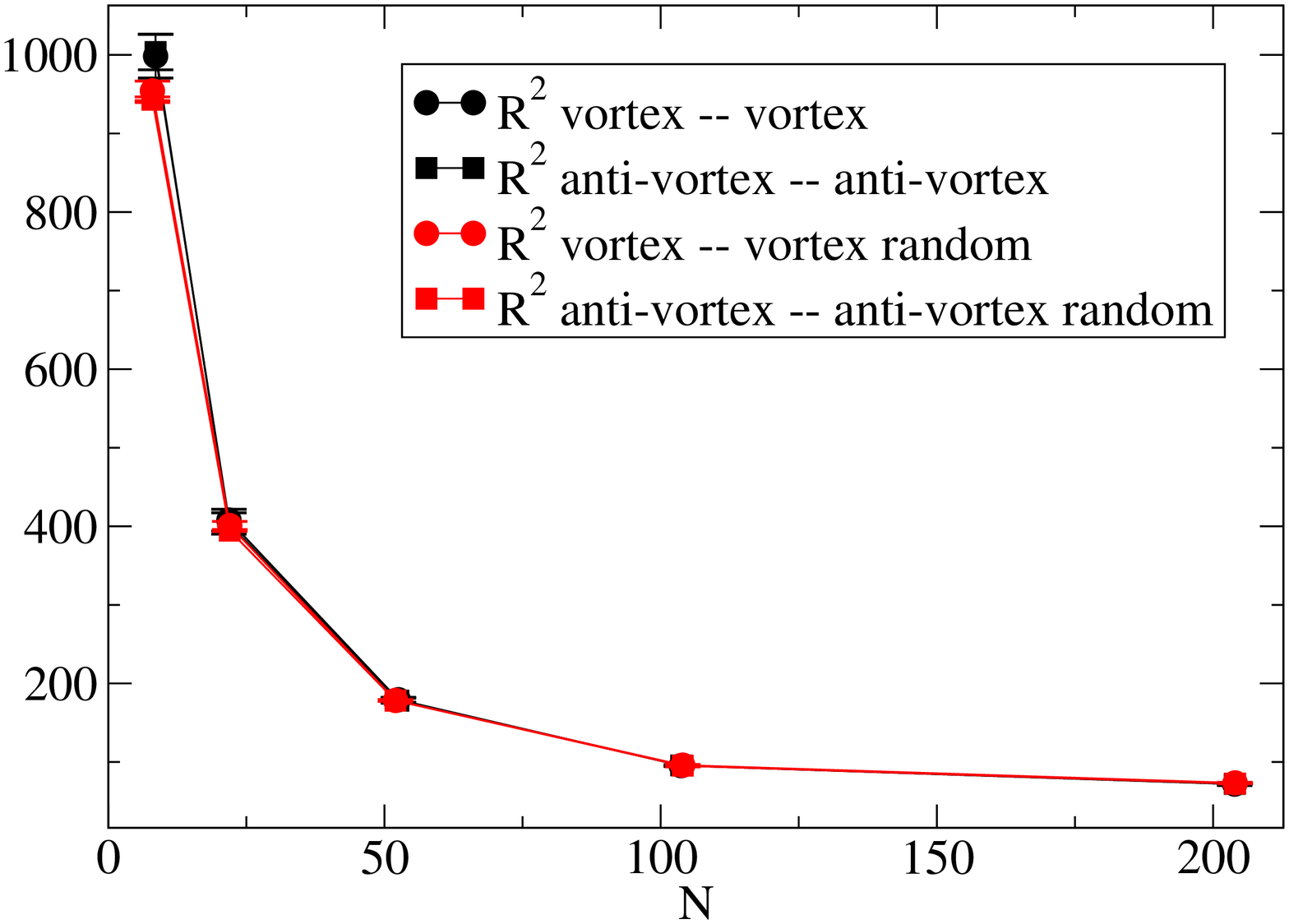}
\vspace*{-5mm}
\caption{\it The quantities $D^{2}$ (above), $R_{\rm VV}^{2}$ and 
$R_{\rm AA}^{2}$ (below) of eqs.\ (\ref{D2}), (\ref{R2}).
In each case, we compare the values for simulated configurations
(at a $\delta$ angle, which leads on average to $N$ vortices)
to $N$ randomly distributed vortices and anti-vortices.
For $D^{2}$ this makes a sizable difference, in particular at low
$\delta$, due to the pair formation. On the other hand, 
such a formation mechanism does not exist for vortex--vortex pairs,
or anti-vortex--anti-vortex pairs, as the results
for $R_{\rm VV}^{2}$ and $R_{\rm AA}^{2}$ reveal.}
\label{D2R2vsran}
\end{figure}

To take a closer look at the meaning of the $D^{2}$ values
in Figure \ref{annealplot}, we add the following comparison:
we take the mean vortex number $N$ at some $\delta$ angle,
and spread the same number of $N$ vortices and anti-vortices
randomly (with a flat probability) over the $L=128$ lattice.
Figure \ref{D2R2vsran} (above) compares the $D^{2}$ 
values obtained in this way to those of the simulation.
We see a large difference, {\it i.e.}\ a strong trend towards
non-accidental pair formation in the simulation, 
in particular up to $N \approx 50$,
which corresponds to $\delta \approx 1.9$ in the simulation;
for larger $N$ this trend is much weaker.

As a further reference quantity, we add (at even $N$) the same 
comparison for
\be \label{R2}
R_{\rm VV}^{2} = \frac{2}{N} \, \sum_{i=1}^{N/2} d_{{\rm VV},\, i}^{\, 2} \quad 
{\rm and} \quad
R_{\rm AA}^{2} = \frac{2}{N} \, \sum_{i=1}^{N/2} d_{{\rm AA},\, i}^{\, 2} \ ,
\ee
where $d_{{\rm VV},\, i}$ ($d_{{\rm AA},\, i}$) are the distances
between two vortices (two anti-vortices), if we perform the 
pairing which minimises $R_{\rm VV}^{2}$ ($R_{\rm AA}^{2}$).
This process is somewhat different from the case of opposite 
pairs (there are only $(N-1)!!$ possibilities). Figure
\ref{D2R2vsran} (below) shows that for these quantities 
(which statistically coincide) it makes hardly any difference
if we take simulated or random distributed vortices (and
anti-vortices). Hence the trend for pair formations
is in fact specific to the pairs of opposite partners.

\begin{figure}[h!]
\centering
\vspace*{-5mm}
\includegraphics[angle=270,width=0.85\linewidth]{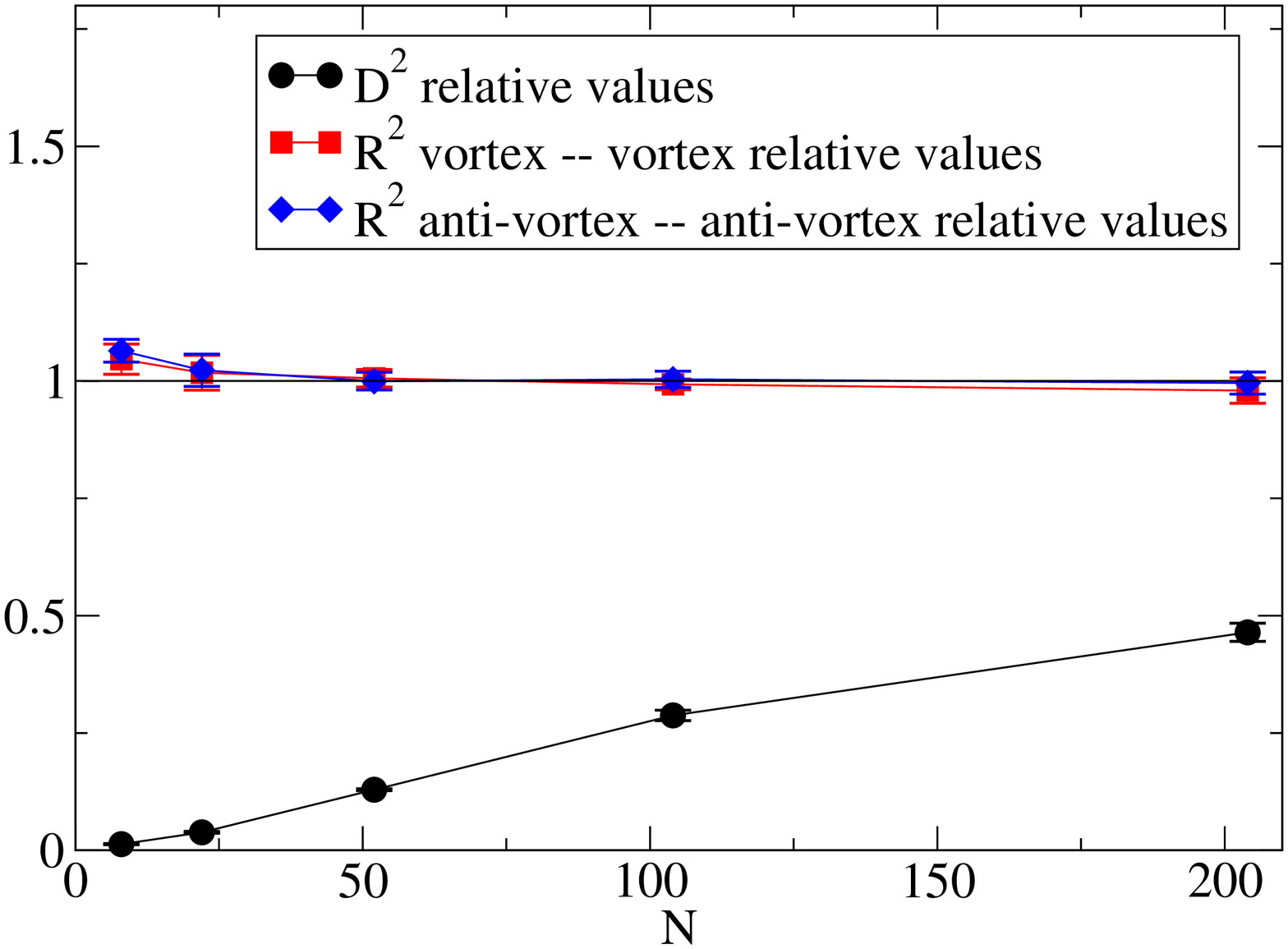}
\vspace*{-5mm}
\caption{\it The ratio between the quantities $D^{2}$, $R_{\rm VV}^{2}$
and  $R_{\rm AA}^{2}$ for simulated configurations, and for random 
distributions, with $N$ vortices and $N$ anti-vortices.}
\label{D2R2vsran2}
\end{figure}

For an ultimate clarification, Figure \ref{D2R2vsran2} shows the 
ratios between $D^{2}$ for simulated configurations with $N$ vortices
and $N$ anti-vortices, and with the same number of randomly
distributed vortices and anti-vortices. We see again
the powerful trend towards pair formation at small $N$, which
fades away as $N$ increases. The plots also shows the corresponding 
ratios for $R_{\rm VV}^{2}$ and for $R_{\rm AA}^{2}$.
In those cases the difference between simulation and random
distribution is tiny; for small $N$ (corresponding to
$\delta \lsim 1.9$) we even observe a slight trend of repulsion
between simulated vortices.

\section{Conclusions}

We have investigated the phase transition of the 2d XY model
in the formulation with the constraint lattice action (\ref{conact}). 
Simulations with dynamical boundary conditions confirmed --- in an
unprecedented manner --- the value of the dimensionless
helicity modulus $\bar \Upsilon_{\rm c,\, theory}$, 
which was predicted for a BKT phase transition.
In contrast to other lattice actions, the finite size effects
are modest in this case. 
In particular, the value of $\bar \Upsilon_{\rm c}$ remains close to
the BKT value, up to $1.9 ~ \%$ down to volumes as small as $16 \times 16$.

This eliminates any doubt that the constraint 
lattice action does belong to the same universality class as the
conventional lattice actions, which involve spin couplings,
like the standard action (\ref{stanact}).
Moreover, this provides one of the most compelling numerical evidences 
that has ever been found for the BKT behaviour of the 2d XY model.

Regarding the spin wave predictions for this model, however,
we observed discrepancies in the range of lattice sizes that
we could explore. This is seen clearly for the coefficient $c_{1}$ 
in the leading finite size correction to $\bar \Upsilon_{\rm c,\, theory}$, 
cf.\ eq.\ (\ref{UpsLeq}), although the universality of $c_{1}$ 
is also supported by renormalization group flow arguments \cite{Peli}.
In this context, however, our data are restricted to $L \leq 256$.

On much larger lattices we still observed small but significant 
deviations from the values  predicted by spin wave theory for
$U_{4}$ and $\xi_{2}/L$. In this case there are also deviations 
for the standard action, at least up to the largest volumes that 
have been simulated. For both quantities, the data obtained for 
the two actions deviate from the spin wave prediction with opposite 
signs.

For the constraint action there is an alternative angle 
$\delta_{\rm c,sw}$, which leads to good agreement with the spin 
wave predictions for $U_{4}$ and $\xi_{2}/L$. However, adopting
that value does still not fix the discrepancy with
the prediction for the coefficient $c_{1}$. In addition it is
not compatible with our data for the correlation length $\xi$,
and this approach to determine the critical point
does not work well for the standard action.

Finally we verified the picture of vortex--anti-vortex pair
(un)binding as the mechanism behind the BKT transition.
Our results for the density of free vortices and anti-vortices 
(without an opposite partner up to some distance), for 
the vorticity correlation, and for the sum over pair separations 
squared, are all compatible with this picture. 
As for the conventional actions, we can indeed 
confirm that a relevant number of pairs dissociate when we
move from the massless to the massive phase, so that the free vortex 
density becomes significant. However, this consideration alone
would not enable a precise determination of the critical 
constraint angle $\delta_{\rm c}$, in analogy to the standard 
action, where the critical value $\beta_{\rm c}$ could only be 
estimated approximately based on the pair (un)binding. In both
cases, the obvious onset of the free vortex density is shifted
somewhat into the massive phase.

The validity of this mechanism is highly non-trivial, since the
free vortices do not cost {\em any} energy (if the constraint
allows them). Hence their suppression in the range
$\pi/2 < \delta \lsim \delta_{\rm c}$ can only be explained
by the combinatorial frequency of configurations carrying different 
vorticities. In this regard, our results deviate from the established 
point of view, since they demonstrate that a BKT transition can occur 
even without any Boltzmann factor suppression of free vortices.

To summarise, the quantitative BKT prediction for the helicity
modulus in the thermodynamic limit is confirmed excellently,
and the picture behind it as well.
On the other hand, the inspiration of this picture --- with a 
Boltzmann weight for free vortices --- is not confirmed, since we 
observe the same feature on purely combinatorial grounds.

\vspace*{6mm} 

{\small
\noindent
{\bf Acknowledgements:} Michael B\"{o}gli has contributed to this
work at an early stage. 
We thank him in particular for providing the data that we used in
Figure \ref{nucritfigs}. We further thank Uwe-Jens Wiese for
instructive discussions, and Martin Hasenbusch and Ulli Wolff for 
interesting remarks.
 
This work was supported by the Mexican {\it Consejo Nacional de Ciencia 
y Tecnolog\'{\i}a} (CONACyT) through project 155905/10 ``F\'{\i}sica 
de Part\'{\i}culas por medio de Simulaciones Num\'{e}ricas'' and 
through the scholarship 312631 for graduate studies, as well as 
DGAPA-UNAM. The simulations were performed on the cluster of the 
Instituto de Ciencias Nucleares, UNAM. We thank Luciano 
D\'{\i}az and Enrique Palacios for technical support.
}

\appendix

\section{Heat bath algorithm}

When we run a Metropolis algorithm for the constraint action
given in eq.\ (\ref{conact}), the decision about accepting
a proposed update step is fully deterministic:
if the updated configuration obeys the constraint, it will always
be accepted, otherwise it must be rejected. This unusual property
holds both for updated spin variables and --- in case of dynamical 
boundary conditions --- also for the updated twist angle $\alpha$.

In this algorithmic scheme it is 
optimal to update the spins one by one. If we suggest a new spin 
\be
\vec e_{x} = \left( \begin{array}{c} \cos \phi_{x} \\ \sin \phi_{x}
\end{array} \right) \to
\vec e_{x}{\, '} = 
\left( \begin{array}{c} \cos \phi_{x}{'} \\ \sin \phi_{x}{'}
\end{array} \right) \ ,
\ee 
it depends on its four nearest neighbours if it will be accepted, 
and --- if $x$ is next to the twisted boundary --- also on $\alpha$. 
The common Metropolis implementation would suggest a new
angle in some small interval around the previous one, $\phi_{x}{'} \in
[ \phi_{x} - \Delta , \phi_{x} + \Delta ]$.

It is noteworthy, however, that for $\delta > \pi/2$ (the case 
we are dealing with) the allowed circular section for 
$\phi_{x}{'}$ may consist of disjoint arcs. Hence by using a
small $\Delta$, one could overlook part of these
allowed arcs, far away from $\phi_{x}$, although those angles
are not suppressed. Therefore it is better to
probe the entire allowed circular section, and choose $\phi_{x}{'}$
therein with flat probability. The situation is similar for the
update of the twist angle $\alpha \to \alpha'$, which should
also be chosen arbitrarily in the allowed circular section 
(which does not violate the constraint). 

Thus the new values do not depend on
the previous ones, but only on (part of) the rest of the 
configuration, so this is a {\em heat bath algorithm.}
This algorithm is robust and generally applicable for this
type of actions. It was also used for quenched QCD simulations 
with an analogous constraint \cite{gaugeconstraint}, {\it i.e.}\ a 
lower bound for each plaquette variable (which was in that case 
combined with a kinetic term).

One could compute the boundaries of the allowed section, including 
possible disjoint pieces, but in particular in the case of $\alpha '$ 
this tends to be tedious. A simpler method suggests
a new angle, $\phi_{x}{'}$ or $\alpha '$, anywhere on the circle
(with flat probability), and checks if it is allowed.
If not, one tries again and repeats this until one finds
an allowed one --- this is a {\em multi-hit procedure.} In practice
one may also limit the number of proposals (``hits'') for the
angle under consideration; if it is still not accepted after
some maximal number of attempts, one moves on to the update 
of another angle.

For spin updates and $\delta \approx \delta_{\rm c}$,
in average 2 proposals for $\phi_{x}{'}$ are sufficient.
More delicate is the search for an acceptable $\alpha '$; here the 
required hit number is much larger and it increases with the lattice
size $L$. At $\delta_{\rm c}$ it takes on average $23.44(4)$ hits 
for $L=16$, $109(4)$ hits for $L=128$, and $162(5)$ hits for $L=256$.
Still, for the study of the helicity modulus the fluctuation
of $\alpha$ is vital, so we have to make sure to choose 
the cutoff for the number of $\alpha '$ hits large enough to 
keep the twist angle moving frequently.

\section{Cluster algorithm}

Sections 3 and 4 deal with periodic boundary conditions,
where it is straightforward and profitable to apply the Wolff 
cluster algorithm \cite{Wolff}. For a given spin configuration 
we choose a random Wolff direction $\vec r$, which defines a {\em spin 
flip} as a reflection on the line through $0$, which is vertical 
to $\vec r$. We then consider pairs of nearest neighbour spins:
if the flip of one of them would violate the
$\delta$-constraint, we connect them by a {\em bond.} Similar to 
Appendix A, we encounter the peculiarity that the bonds are set in 
a fully deterministic way. A set of spins connected by bonds constitutes 
one {\em cluster.} Thus we can divide the whole lattice into 
clusters (sets of spins, which can only be flipped collectively)
and flip each one with probability $1/2$ 
{\em (multi-cluster algorithm)}, or we start from a random site and 
build one cluster,
which will be flipped for sure {\em (single-cluster algorithm).}
In either case, it is guaranteed that the new configuration obeys
the $\delta$-constraint, since the collective flip of two spins
in the same cluster just changes the sign in their relative angle,
\be
\Delta \phi_{x,y} 
\, \rightarrow  \, - \Delta \phi_{x,y} \ .
\ee
This algorithm is far more efficient than the heat bath method,
in particular close to $\delta_{\rm c}$.\\

Hence it is highly motivated to search for a generalised cluster algorithm, 
which can also be applied in the presence of {\em dynamical boundary 
conditions,} that we are confronted with in Section 2.
This issue has been addressed before in Ref.\ \cite{Ols}, for a
setting where the twist is split over the layers of lattice sites, 
though that algorithm does not apply to all configurations.

In our setting, the difficulty can be seen as follows.
Assume that we build clusters of spins 
by implementing the instruction to put a bond whenever the flip
of one spin out of a nearest neighbour pair would be forbidden.
However, the actual goal behind this instruction
is that clusters can be flipped freely without ever violating
the constraint. If we flip a cluster which includes a spin
pair across the twisted boundary, its relative angle\footnote{The 
modulo operation is still defined such that it provides
a minimal absolute value, cf.\ eq.\ (\ref{DeltaPhi}).} 
\be
\Delta \phi_{x_{2}}^{\rm tb} = ( \phi_{L, x_{2}} - \phi_{1, x_{2}})
\ {\rm mod} \ 2\pi \ , 
\quad x_{2} \in \{ 1, \dots, L\} 
\ee
before the flip obeys 
\be  \label{alphacond}
| \, (\Delta \phi_{x_{2}}^{\rm tb} + \alpha ) \ {\rm mod} \ 2 \pi \, | 
< \delta \ .
\ee
A cluster flip (in the above sense) only entails the transition 
$\Delta \phi_{x_{2}}^{\rm tb} \to - \Delta \phi_{x_{2}}^{\rm tb}$, so it 
is {\em not} guaranteed anymore that the constraint still holds.

We could guarantee that for the cluster under consideration if
we flip simultaneously the sign of $\alpha$. However, if the
cluster does not capture the entire twisted boundary, this
sign change could lead to a violation of the constraint for
other spin pairs.
Hence we are forced to include $\alpha$ in the cluster building. 
The inclusion of such a non-local variable --- which extends in this 
case over a whole boundary --- would be a conceptual novelty.

\subsection{Proposal for a cluster algorithm with dynamical
boundary conditions}

Here we sketch an attempt to construct a cluster algorithm that
involves the twist angle $\alpha$, and its shortcomings.
We still assume only one twist, at the boundary between $x_{1}=1$ 
and $L$. Then the constraint across this boundary takes the form
(\ref{alphacond}).



We stay with the prescription that a flip of a spin changes the 
sign of its component parallel to the Wolff direction, while
flipping $\alpha$ means
\be  \label{alphaflip}
\alpha \rightarrow - \alpha \ .
\ee
Cluster updates alone are not ergodic, since they never change 
$| \alpha |$. Hence one has to insert intermediate steps which do 
perform such a change; this is done best by a heat bath 
$\alpha$-update, as described in Appendix A.\\

However, the real issue is {\em the formation of the clusters,} 
such that they can be flipped freely while Detailed Balance is 
guaranteed. The spin part of a cluster grows in the usual way.
Once a cluster touches the twisted boundary, {\it i.e.}\ once 
it incorporates a spin variable $\vec e_{1,x_{2}}$ or $\vec e_{L,x_{2}}$, 
it is possible to add $\alpha$ and/or the periodic neighbour 
spin to this cluster. If this happens for $\alpha$, any spin 
$\vec e_{1,x_{2}{'}}$ or $\vec e_{L,x_{2}{'}}$ (with $x_{2}{'} \neq x_{2}$)
could further join the cluster. Thus a ``cluster'' could
consist of disconnected patches. 

Let us consider two spins $\vec e_{1,x_{2}}$ and $\vec e_{L,x_{2}}$
(for some $x_{2}$) and the twist angle $\alpha$. 
We discuss the options to put a bond which ties 2 of these
3 variables, or even a super-bond which captures all the 3,
so they all belong to the same cluster. 
In light of the above flip definition, there are 
{\em 8 possible constellations:} the projection of the spins
in the Wolff direction, and the angle $\alpha$, can all be
positive or negative. We have also pointed out before that a 
collective flip of all 3 variables cannot violate the 
constraint.\footnote{Of course, we assume the system to be 
initially in an allowed configuration.}
Hence we restrict our table to the case $\alpha > 0$; the rest 
is determined by the invariance under a collective flip. 
We distinguish the orientations of the two spins
($\upa$ or $\doa$ with respect to the Wolff direction) at 
$\alpha > 0$. For each of these 4 options, the action can be
$0$ or $+ \infty$, so there are $16$ cases, see Table \ref{bondtab}. 

In each case we specify which among the 3 variables have to be 
tied by a bond in order to exclude cluster flips that could 
violate the constraint.
These variables are given as the indices of the bond term $B$, where 
$1$ and $L$ refer to the spins $\vec e_{1,x_{2}}$ and $\vec e_{L,x_{2}}$, 
respectively. 
Also these bonds are set in a fully deterministic manner.

\begin{table}[h!]
\centering
\begin{tabular}{|c|c||c|c|c|c|c|c|c|c|}
\hline
$\vec e_{1,x_{2}}$ & $\vec e_{L,x_{2}}$ & \multicolumn{8}{|c|}{$\exp (-S)$} \\
\hline
\hline
$\upa$ & $\upa$ & $1$ & $1$ & $1$ & $1$ & $0$ & $1$ & $1$ & $0$ \\
\hline
$\upa$ & $\doa$ & $1$ & $1$ & $1$& $0$ & $1$ & $1$ & $0$ & $1$ \\
\hline
$\doa$ & $\upa$ & $1$ & $1$ & $0$ & $1$ & $1$ & $0$ & $1$ & $1$ \\
\hline
$\doa$ & $\doa$ & $1$ & $0$ & $1$ & $1$ & $1$ & $0$ & $0$ & $0$ \\
\hline
\hline
\multicolumn{2}{|c||}{case number} & 1 & 2 & 3 & 4 & 5 & 6 & 7 & 8 \\
\hline
\multicolumn{2}{|c||}{bond} & --- & $B_{1L \alpha}$ & $B_{1L \alpha}$ 
& $B_{1L \alpha}$ & $B_{1L \alpha}$ & $B_{1 \alpha}$ & $B_{L \alpha}$ 
& $B_{1 L}$ \\
\hline
\end{tabular}

\vspace*{7mm}

\begin{tabular}{|c|c||c|c|c|c|c|c|c|c|}
\hline
$\vec e_{1,x_{2}}$ & $\vec e_{L,x_{2}}$ & \multicolumn{8}{|c|}{$\exp (-S)$} \\
\hline
\hline
$\upa$ & $\upa$ & $1$ & $0$ & $0$ & $1$ & $0$ & $0$ & $0$ & $0$ \\
\hline
$\upa$ & $\doa$ & $0$ & $1$ & $0$ & $0$ & $1$ & $0$ & $0$ & $0$ \\
\hline
$\doa$ & $\upa$ & $0$ & $0$ & $1$ & $0$ & $0$ & $1$ & $0$ & $0$ \\
\hline
$\doa$ & $\doa$ & $1$ & $1$ & $1$ & $0$ & $0$ & $0$ & $1$ & $0$ \\
\hline
\hline
\multicolumn{2}{|c||}{case number} & 9 & 10 & 11 & 12 & 13 & 14 & 15 & 16 \\
\hline
\multicolumn{2}{|c||}{bond} & $B_{1L}$ & $B_{L\alpha}$ & $B_{1\alpha}$ 
& $B_{1L\alpha}$ & $B_{1L\alpha}$ & $B_{1L\alpha}$ & $B_{1L\alpha}$
& $\exists \!\!\! /$ \\
\hline
\end{tabular}
\caption{\it Table for the orientations of two spins at the twisted 
boundary. We assume $\alpha > 0$, consider the possible actions
in each case, and specify which bonds are necessary to avoid flips to
forbidden configurations.}
\label{bondtab}
\end{table}

\begin{itemize}

\item Two cases in Table \ref{bondtab} are trivial:
case 1 (no bond, so flips lead to all 8 constellations) 
and case 16 (never occurs). 

\item In cases $12 \dots 15$, 3 spin orientations are forbidden.
Here a super-bond over all three variables is required,
which just allows for a total flip of both spins and $\alpha$,
thus including exactly the 2 allowed constellations.

\item In cases $6 \dots 11$, there are two forbidden spin orientations.
In these cases, we 
put one bond which ties
2 out of the 3 variables; then going through all
flips covers exactly the 4 allowed constellations.

\item In cases $2 \dots 5$ there is one forbidden spin orientation,
and excluding it requires again a super-bond over all 3 variables. 

Unlike the previous cases, this is not really consistent, 
because this super-bond only allows for flips between 2 
constellations of the spins and of $\alpha$, missing the other 
4 allowed constellations. Actually it is obvious that we cannot 
get access to all the 6 allowed constellations in this way,
because flips can only attain to $2^{n}$ constellations,
$n \in \N$.

\end{itemize}

Not allowing a transition to the missing 4 constellations in cases
$2 \dots 5$ implies a violation of Detailed Balance, so this is 
a serious problem (similar to the approach of Ref.\ \cite{Ols}).
A brute force solution is to put {\em no} bond in these cases, and 
suggest cluster flips, which are rejected if they violate the constraint.

To probe if this mixed approach is promising, we investigate 
how often these troublesome constellations occur. 
There are 5 angles involved: the spin angles 
$\phi_{1,x_{2}}$ and $\phi_{L,x_{2}}$, the twist angle $\alpha$, 
the angle of the Wolff direction and the constraint
angle $\delta$.
Again we assume $\alpha \geq 0$, we fix $\delta = \delta_{\rm c}$
and we now distinguish the {\em classes} with $k = 1$, 2, 3 or 4 
allowed spin orientations. 
The ``bad cases'' are those in the class $k=3$ (cases $2 \dots 5$ 
in Table \ref{bondtab}); one might hope that they are rare. 

\begin{figure}[h!]
\centering
\includegraphics[angle=270,width=0.8\linewidth]{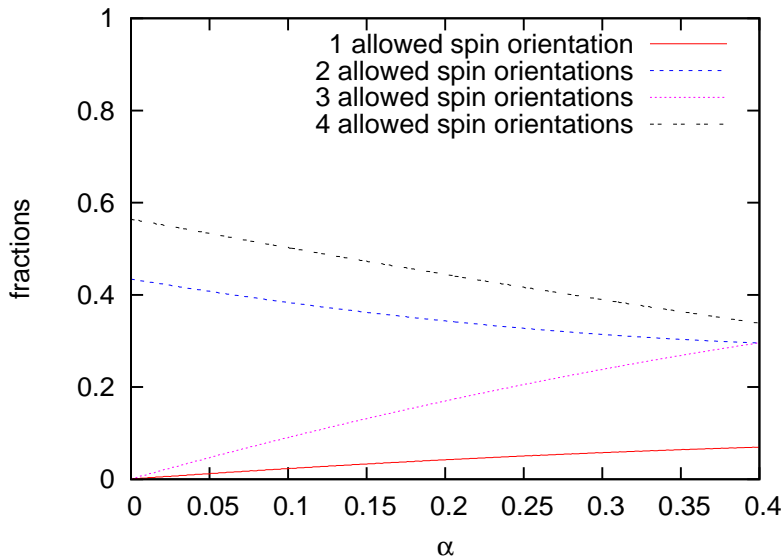}
\caption{\it The relative frequency of the classes with $k=1$, $2$, 
$3$ or $4$ allowed spin orientations, at $\delta = \delta_{\rm c}
= 1.77521$, for variable angles of the Wolff direction and the 
two spins (with an allowed starting orientation).}
\label{fractplot}
\end{figure}

We consider $\alpha \in [0, \, 0.4]$. At fixed $\alpha$, we
vary the 3 remaining angles (Wolff direction and the 
two spins, such that we obtain an allowed spin orientation),
and count which fraction belong to each of the 4 classes.
This is shown in Figure \ref{fractplot}. At $\alpha = 0$
the ``odd classes'' ($k=1$ or 3) do not occur. When we keep
$\alpha$ small, they are still suppressed.
This suppression is strong for $k=1$, but not that much 
for $k=3$, unfortunately.
For instance, at $\alpha =0.1$ already $9 \ \%$ of the angular
combinations belong to class $k=3$.

The brute-force method is not a consistent cluster algorithm;
this could be harmful for its efficiency. In view of its
occasional accept/reject decision about cluster flips, and the 
additional heat bath step (which is required for the alteration
of $|\alpha |$), this algorithm does not appear very promising.
Hence we did not apply it, 
and it remains an interesting open question how to deal with 
the cases in the class $k=3$, and --- more generally --- how to 
construct a consistent cluster algorithm which involves a non-local 
variable, such as a dynamical boundary twist.

\section{Simulated annealing}

For a given configuration, we first identify the
$N$ vortices and $N$ anti-vortices. There are $N!$ 
possibilities to build vortex--anti-vortex pairs. For one 
option, they are separated by some Euclidean distances 
$d_{{\rm VA},\, i}$, $i = 1 \dots N$.
For the considerations in Subsection 4.3 we search for the 
pairing which minimises the term $D^{2}$ in eq.\ (\ref{D2}). 
For comparison we also perform the optimal pairing among the
vortices and among the anti-vortices. On the $L=128$  
lattice, and $\delta$ close to --- or above --- $\delta_{\rm c}$,
we cannot check all possibilities, since $N$ is of $O(100)$.\footnote{On 
one core at 2.9 GHz, that takes almost 7 hours for $N=14$, 
and more than 4 days for $N=15$.} Hence we resort to the 
technique of {\em simulated annealing} \cite{anneal}.

We start from one arbitrary pairing, measure $D^{2}$
and suggest a minimal modification by exchanging the partners
among two pairs (which are randomly selected). 
Then we take a Metropolis-style decision about this modification: 
it is always accepted if $D^{2}_{\rm new} \leq D^{2}_{\rm old}$, 
and with probability $\exp ( [D^{2}_{\rm old} - D^{2}_{\rm new}]/T )$ otherwise.
The parameter $T$ decreases as a monotonous annealing function of 
the number of $\tau$ of such annealing steps. 
We test 3 functions of this kind: linear, exponential and piecewise 
constant,
\be
T (\tau ) = \left\{ \begin{array}{ccc}
T_{0}(1 - \tau / u ) && \tau / u < 1 \\
0 && {\rm otherwise} \end{array} \right. \ , \quad
T(\tau ) = T_{0} \cdot v^{1000 \tau /u } \ ,
\ee
or the deformation of the linear decrease into stair steps.
The constants are chosen as $T_{0} = 100$, $u= 10^{7}$ and $v=0.99$.

For a given configuration, we test $20$ initial pairings, 
all 3 annealing functions, and the process ends based 
on a ``Cauchy criterion'' (no change within 500 annealing steps).
The lowest final $D^{2}$ value found in this way is in fact 
the global minimum for configurations with up to 14 vortices,
as we checked by testing all pair formations. For higher vortex 
numbers we consider this way of searching for
the global minimum still quite reliable, based on the 
consistency of the optimal solution identified in multiple
runs, with distinct starting points and annealing functions.

\end{document}